\documentclass{ieeeaccess}
\usepackage{amsmath,amssymb,amsfonts}
\usepackage{algorithmic}
\usepackage{multirow}
\usepackage{graphicx}
\usepackage{caption}
\usepackage{booktabs}
\usepackage{makecell, multirow, rotating, graphicx}
\usepackage[ruled,lined,linesnumbered]{algorithm2e}
\usepackage{float}
\usepackage{placeins}

\usepackage{textcomp}
\def\BibTeX{{\rm B\kern-.05em{\sc i\kern-.025em b}\kern-.08em
    T\kern-.1667em\lower.7ex\hbox{E}\kern-.125emX}}
\usepackage[
  backend=biber,
  style=ieee,
  maxnames=5,
  minnames=3,
  doi=false,
  isbn=false,
  url=false,
  eprint=false
]{biblatex}
\usepackage[hidelinks]{hyperref} 
\usepackage{xcolor}
\usepackage{soul}
\sethlcolor{yellow}
\usepackage{empheq}
\begin{document}

\title{Decomposition-Based QAOA for Maximum Coverage Location Problem in Satellite Constellation Design}
\author{\uppercase{Divya Sisodiya}\authorrefmark{1},
\uppercase{Amiratabak Bahengam\authorrefmark{2}, Hang Woon Lee}\authorrefmark{3}, 
\IEEEmembership{Member, IEEE}, AND \uppercase{Hao Chen}\authorrefmark{2}, \IEEEmembership{Member, IEEE}}
\address[1]{Department of Physics, Stevens Institute of Technology, Hoboken, NJ 07030 USA}
\address[2]{Department of Systems Engineering, Stevens Institute of Technology, Hoboken, NJ 07030 USA}
\address[3]{Department of Mechanical, Materials and Aerospace Engineering, West Virginia University, Morgantown, WV 26506 USA}

\corresp{Corresponding author: Hao Chen (e-mail: hao.chen@stevens.edu).}

\begin{abstract}
An increase in earth observation missions has increased the demand of efficient design and optimization of satellite constellations. Maximizing coverage of the target while effectively utilizing the limited orbital resources is one of the critical design challenges for complex combinatorial optimization problems. The maximal covering location problem (MCLP), serves as a base for orbital coverage modeling, is NP-hard and computationally intractable for large-constellation instances. Using heuristics, metaheuristics, and mixed-integer linear programming, classical solvers have achieved optimal or near-optimal results, yet their scalability is limited as the problem size increases. Quantum computing advancements, including the quantum approximate optimization algorithms, offer a potential solution to NP-hard combinatorial optimization problems. Current quantum hardware limitations, such as low qubit counts and circuit depth, restrict solutions for small-scale instance problems. To address this challenge, this paper proposes a scalable quantum optimization framework for MCLP in satellite constellation design. A decomposition-based quantum methodology is proposed, in which large MCLP instances are partitioned into subgraphs by classical decomposition, optimized independently via quantum optimization circuits, and combined using quantum reconstruction strategies. Computational results across different constellation sizes reveal better scalability in less time while maintaining competitive coverage performance compared to classical solvers. 

\end{abstract}


\titlepgskip=-15pt

\maketitle

\section{Introduction}
\label{sec:introduction}
Satellite constellations address fundamental limitations of single-satellite Earth observation systems. A single satellite in low earth orbit can only observe a given ground location during brief overpass windows, typically lasting minutes every several days. These long temporal gaps create operational challenges for time-sensitive applications such as disaster response, environmental monitoring, and intelligence gathering, where critical monitoring is needed. Optimal satellite constellations can also address resource constraints by strategically placing a fixed number of satellites across orbital slots to minimize revisit times and maximize temporal coverage. NASA and commercial entities drive significant investment to provide reduced revisit times and enhance operational redundancy compared to single-satellite missions. Nevertheless, the combinatorial complexity of selecting orbital slots for coverage maximization poses substantial optimization challenges.

While multiple formulation frameworks exist for constellation design, including the set covering location problem (SCLP), partial set covering location problem (PSCLP), and maximum covering location problem (MCLP), the MCLP framework is particularly well-suited for the resource-constrained satellite constellation design problem. The significance of MCLP applied to the constellation configuration design optimization is that it enables the user to design a constellation where the number of satellites to be used is a parameter and the objective is to obtain the maximum observational rewards over a set of targets \cite{rogers_optimal_2026}. This objective is distinctive from those of SCLP and PSCLP, which seek to provide continuous coverage or a percentage of coverage over the targets while considering the number of satellites as a decision variable instead of a parameter \cite{rogers_optimal_2026}. Originally, it was developed by Church and ReVelle, MCLP formulation maps orbital slots to facility locations, represents satellites as limited resources, and ground targets as demand nodes requiring coverage \cite{church_maximal_1974}.

Previous work on the satellite constellation optimization included complex mathematical and heuristic methodologies of several types. The optimal configuration of the satellite constellation is modeled as a multi-objective mixed integer programming problem, and a metaheuristic optimization method is applied to optimize the configuration and number of satellite ground tracks and orbital planes \cite{imoto_optimal_2023}. An integer linear programming (ILP) based framework for common ground track constraints reduces the worst-case revisit time by over 60\% compared to a symmetric constellation \cite{ko_satellite_2026}. A unifying framework is proposed consisting of five mixed-integer linear program formulations of practical significance, extensible to more complex mission narratives using additional constraints, and capable of obtaining the optimal constellation configurations \cite{rogers_optimal_2026}. Satellite reconfiguration has been made as a novel bi-objective ILP formulation that combines constellation design and transfer problems \cite{lee_regional_2023}.  MCLP is extended through a solution procedure for a probabilistic model for one server per center, using a column generation and covering graph approach, where the computational tests report results for network instances up to 818 vertices \cite{correa_decomposition_2009}.  Despite these optimization strategies demonstrating strong performance, some key limitations persist. While meta-heuristic approaches such as ant-colony optimization have been successfully applied to constellation design problems, their scalability to large instances remains unvalidated \cite{imoto_optimal_2023}. Efficient computing algorithms are required for constellations to optimize higher resolution design problems such as decomposition methods and heuristic-guided solution algorithms for the
ILP, which are scalable for complicated instances \cite{ko_satellite_2026}. Existing mixed-integer linear programming (MILP) formulations for constellation design provide provable optimality guarantees for discrete orbit sets but face limitations with computational runtime that hinder scalability to larger instances \cite{rogers_optimal_2026}. Classical solution methods, including MILP solvers, Lagrangian relaxation heuristics, and branch-and-bound algorithms, face computational intractability for large-scale instances \cite{lee_regional_2023}. To optimize large-scale constellations, the column-generation method needs further study to remove unproductive columns for reducing the computational time \cite{correa_decomposition_2009}. These limitations highlight the necessity for methods that maintain solution quality with exponential increases, motivating the need for alternative optimization approaches.

 The computational intractability of the classical optimization methods for combinatorial optimization problems has motivated the exploration of alternative methods that are capable of exploring quantum mechanical phenomena for enhanced solution strategies. Quantum computing (QC) leverages superposition and entanglement to enable parallel exploration of solution spaces, which offers potential advantages in addressing NP-hard combinatorial optimization problems that are computationally costly for classical algorithms. Several studies have demonstrated the application of QC techniques to combinatorial optimization, showing enhanced data-driven modeling through augmented representation and optimization \cite{hong_review_2025}, approximate solution generation for the complex problem instances \cite{farhi_quantum_2014}, and successful implementation on MCLP \cite{giraldo-quintero_using_2022}. The quantum approximate optimization algorithm (QAOA), as a hybrid quantum-classical variational approach, employs parametrized quantum circuits to encode problems as Hamiltonians and iteratively refines solutions through classical parameter optimization \cite{farhi_quantum_2014}. QC is applied to solve the maximal covering location problem proposed by Church and ReVelle \cite{giraldo-quintero_using_2022}. QAOA has shown promise in solving combinatorial optimization problems by using quantum computational power \cite{minato_two-step_2025}. Two-step QAOA can improve the effectiveness of QAOA by decomposing problems with k-hot encoding QUBO (Quadratic Unconstrained Binary Optimization) formulations \cite{minato_two-step_2025}. Quantum solvers such as D-Wave quantum annealers have demonstrated potential in optimization by formulating problems as QUBO, with results showing improvements in computational efficiency and solution quality over classical methods \cite{marchioli_quantum_2025}. Even though quantum optimization can be computationally efficient and yield promising results, its scalability is limited due to qubit count, connectivity, and gate fidelity. Applications are limited to small instances on D-Wave and IBM devices, as large-scale instances cannot be solved due to constraints in qubit count and circuit depth \cite{hong_review_2025,farhi_quantum_2014,giraldo-quintero_using_2022}.  Even advanced methods of quantum optimization methods face challenges in addressing more complex instances, suggesting that other quantum or quantum-inspired approaches could be explored to evaluate their solutions \cite{minato_two-step_2025, marchioli_quantum_2025}.

To overcome the limitation of large-scale optimization, divide-and-conquer algorithms have emerged as a widely adopted strategy to split the complicated problems into tractable subproblems that can be solved independently for combining sub-solutions to build one global solution. Most of the existing decomposition algorithms are a product of this algorithm technique \cite{esau_taiwo_comparative_2020}. Several studies adopted this algorithm with quantum algorithms to increase the scalability with in the optimization. Divide-and-conquer QAOA (DC-QAOA) uses graph partitioning methods, such as spectral bisection, to partition the large graphs into subgraphs and reconstruct the global solution using quantum state reconstruction (QSR) \cite{li_large-scale_2023}. The QAOA-in-QAOA (QAOA²) framework uses the $\mathbb{Z}_2$ symmetry to reformulate the merging step as a secondary max-cut problem, resulting in competitive performance on a graph with 2000 nodes using 10-qubit simulators \cite{zhou_qaoa--qaoa_2023}. Decomposition methods using graphs enable a 100-vertex max-cut problem due to a reduction in the qubit requirement by nearly 90\% on the quantum trapped-ion computer \cite{ponce_graph_2025}. Similar instances with more than a thousand variables are effectively handled by problem-specific decomposition for number partitioning on D-Wave \cite{li_efficient_2025}. These methods substantially broaden the applicability of quantum optimization to larger problem instances. At the parameter-optimization level, complementary advances such as tensor-train-guided hypernetworks ~\cite{qi_tensorhyper-vqc_2026}, address noise resilience and training stability, though their application to coverage optimization problems requires further problem-specific adaptation.

Divide-and-conquer quantum methods have been applied to generic combinatorial problems such as graph max-cut and number partitioning. Their application to the MCLP  for satellite constellation design remains unexplored. Addressing this gap, we present a decomposition framework that integrates MCLP formulation with  a divide-and-conquer-based strategy using QC for satellite constellation design. The proposed approach defines a decomposition-based quantum optimization algorithm for MCLP capable of handling large-scale instances. The MCLP is formulated as a graph using a matrix structure, and then its partitioned into subgraphs with a minimum cut objective using a classical decomposition method. Each subgraph is solved independently using QAOA on quantum circuits. The solutions from subgraphs are merged using two approaches applied to the separator nodes: the first integrates the measurements of shared separator nodes to reconstruct the sub-solutions, while the second leverages the graph structure of separator nodes without requiring measurement copies. This framework provides a scalable quantum optimization method for satellite constellations while also enabling comparative analysis of merging techniques for constructing global solutions from subgraph computations.

Our research develops a decomposition-based quantum optimization framework with three principal contributions. First, satellite constellation design is formulated as an MCLP and mapped into a QUBO representation with an Ising Hamiltonian, enabling its solution using quantum optimization techniques. Unlike graph-partitioning problems such as Max-Cut, the MCLP exhibits coverage-demand coupling and cardinality constraints, necessitating a problem-specific formulation and reconstruction strategy. Second, a divide-and-conquer framework is developed that combines classical spectral graph bisection for co-observation graph decomposition, QAOA for subproblem optimization, and two different  Hamiltonian-based reconstruction methods.  Finally, the proposed framework is evaluated across different MCLP instances of varying sizes and compared with Gurobi, and standard QAOA. The results demonstrate that how decomposition enables the treatment of large-scale instances containing hundreds of orbital slots compared to the standard QAOA which faces problem of exponential state-space growth. Validation using IBM Quantum hardware and classical simulation further highlights the practicality of the proposed framework for large-scale satellite constellation design under current NISQ constraints.

The remainder of the paper is organized as follows. Section II introduces the problem formulation consisting of the system model, decision variables, objective function, constraints, and mapping of MCLP to the satellite constellation design problem. Section III presents the proposed decomposition-based methodology, encompassing a framework overview, co-observation graph construction, decomposition and budget allocation strategy, subproblem solver, merging techniques, and algorithmic descriptions. Section IV details the experimental setup that defines problem instances, solver configurations, and performance metrics. Section V presents results and discussion. Section VI provides a conclusion and directions for future work.

\section{Problem formulation
}
\label{sec:problem_formulation}
This section presents the MCLP formulation for satellite constellation design. Section~\ref{sec:system_model} describes the system model and input parameters. Section~\ref{sec:decision variables} defines the decision variables governing satellite placement and target coverage. Section~\ref{sec:obj_function} establishes cardinality and coverage constraints that characterize the feasible constellation configurations. Lastly, section~\ref{sec:constraints} formulates the objective function to maximize the coverage.

\subsection{System Model and Inputs}
\label{sec:system_model}
The satellite constellation comprises a set of orbital slots, each capable of hosting at most one satellite. Satellites deployed on these orbital slots provide coverage to ground targets distributed across the observation region. The system model is defined by the following input parameters:

\textbf{Orbital Slots ($J$):} discrete positions within available orbits where satellites may be deployed.

\textbf{Satellites ($N$):} number of satellites available for deployment on orbital slots. 

\textbf{Ground Targets ($P$):} geographical locations on the earth's surface requiring satellite coverage.

\textbf{Rewards ($\pi_{tp}$):}  non-negative numbers representing the operational value of covering the target $p$ at timestep $t$, defined as $\pi_{tp}\in\mathbb{R}_{\geq 0}$.

\textbf{Time Horizon ($T$):}  the total number of discrete timesteps over which target coverage is evaluated. The time horizon is temporally discretized to balance computational complexity with coverage accuracy.

\textbf{Visibility Matrix $(\Phi_{tjp})$:} a binary tensor indicating whether the target $p$ is visible from the orbital slot $j$ at timestep $t$, defined as $\Phi \in \{0, 1\}^{T \times J \times P}$, where 1 denotes the target $p$ is visible from the orbital slot $j$ at timestep $t$, and 0 denotes the target is not within the coverage region.

These input parameters are subsequently transformed into decision variables, constraints, and an objective function that define the satellite constellation optimization problem.

\subsection{Decision Variables}
\label{sec:decision variables}
The MCLP employs two classes of binary decision variables to represent satellite deployment and target coverage status: 

\textbf{Orbital Slot Selection Variable:} a binary variable $x_j$  indicates the deployment status of orbital slot $j$, and enables direct enforcement of the satellite budget constraint through linear summation, defined as 
\begin{equation}
x_j =
\begin{cases}
1, & \text{if orbital slot } j \text{ is occupied by a satellite} \\
0, & \text{otherwise}
\end{cases}
\label{eq:orbital_slot}
\end{equation}

\textbf{Coverage Variable:} a binary variable $y_{tp}$ indicates whether the target $p$ is covered at time step $t$. It enables efficient aggregation of coverage rewards and provides a linear reformulation of the coverage condition, defined as

\begin{equation}
y_{tp} =
\begin{cases}
1, & \text{if target } p \text{ is covered at time step } t \\
0, & \text{otherwise}
\end{cases}
\label{eq:CoverageVariable}
\end{equation}

Further subsections define constraints and the objective function that govern the relationship between these decision variables.

\subsection{Objective Function}
\label{sec:obj_function}
The objective function maximizes the total weighted coverage of ground targets over the time horizon. Given the binary variable $x_j$, and the coverage indicator $y_{tp}$, the objective function is formulated as follows:
\begin{equation}
\max_{x,y} \quad F = \sum_{t=1}^{T} \sum_{p=1}^{P} \pi_{tp}  y_{tp}
\label{eq:objective}
\end{equation}

Objective function \eqref{eq:objective} aggregates coverage rewards across all the targets and timesteps.  It represents the fundamental objective of MCLP, maximizing total covered demand subject to limited facility deployment.

\subsection{Constraints}
\label{sec:constraints}
The MCLP formulation is subject to two principal constraints governing satellite deployment and target coverage.

\textbf{Cardinality Constraint:} 
 the fixed number of satellites available for constellation deployment, expressed as
\begin{equation} 
\sum_{j=1}^{J} x_j = N
\label{eq:cardinality}
\end{equation}
The cardinality restricts the total number of deployed satellites to exactly $N$. 

\textbf{Coverage Constraint:} a target $p$ is considered covered at a timestep $t$ if at least one deployed satellite maintains visibility to target $p$. This coverage condition is enforced through the linear constraint, expressed as:

\begin{equation}
\sum_{j=1}^{J} \Phi_{tjp} x_j \geq y_{tp}
\quad
\forall\, t \in \{1, \ldots, T\}, \, p \in \{1, \ldots, P\}
\label{eq:coverage_condition}
\end{equation}

The optimization problem defined by the decision variables \eqref{eq:orbital_slot}, \eqref{eq:CoverageVariable}, objective function \eqref{eq:objective}, cardinality constraint \eqref{eq:cardinality}, and coverage constraint \eqref{eq:coverage_condition} constitutes a mixed-integer program belonging to the NP-hard class  of combinatorial optimization problems \cite{church_maximal_1974}.

Unlike graph partitioning problems (e.g., Max-Cut), where the objective depends solely on graph edges and adjacency relationships, the MCLP exhibits a fundamentally different structure. It couples location selection variables $x_j$ with coverage–demand relationships encoded in the visibility tensor $\mathbf{\Phi}_{tjp}$, which indicates whether orbital slot $j$ can observe target $p$ at timestep $t$, and timestep-dependent priorities $\pi_{tp}$. This introduces two key constraints absent in standard graph-based formulations: (1) \textit{Coverage semantics}: selected orbital slots must collectively provide sufficient observation coverage for target at maximum timesteps; and (2) \textit{Cardinality constraint}: exactly $N$ orbital slots must be selected, corresponding to the fixed satellite budget. 

These structural properties necessitate problem-specific algorithmic design, including a coverage-aware Hamiltonian formulation, an MCLP-preserving decomposition strategy, and coverage-compatible reconstruction operations. In contrast, generic decomposition-QAOA frameworks do not directly capture the coverage-dependent coupling between $\mathbf{\Phi}_{tjp}$, $x_j$, and $\pi_{tp}$ inherent in MCLP.

\section{Decomposition-based QAOA Methodology}
\label{sec:methodology}
This section presents a decomposition-based QAOA framework for solving large-scale MCLP instances under qubit constraints. Section III-A provides an overview of the recursive decomposition strategy. Section III-B defines the co-observation graph with nodes and edge weights capturing coverage interactions. Section III-C introduces spectral bisection for graph partitioning via minimum cut optimization. Section III-D presents the satellite allocation strategy across subgraphs. Section III-E describes the quantum algorithm for solving MCLP subproblems. Section III-F introduces quantum merging strategies for constructing global constellations. Section III-G summarizes the complete algorithm.

\subsection{Framework Overview}
\label{sec:framework overview}
\begin{figure}[H]
\centering
\includegraphics[width=\columnwidth]{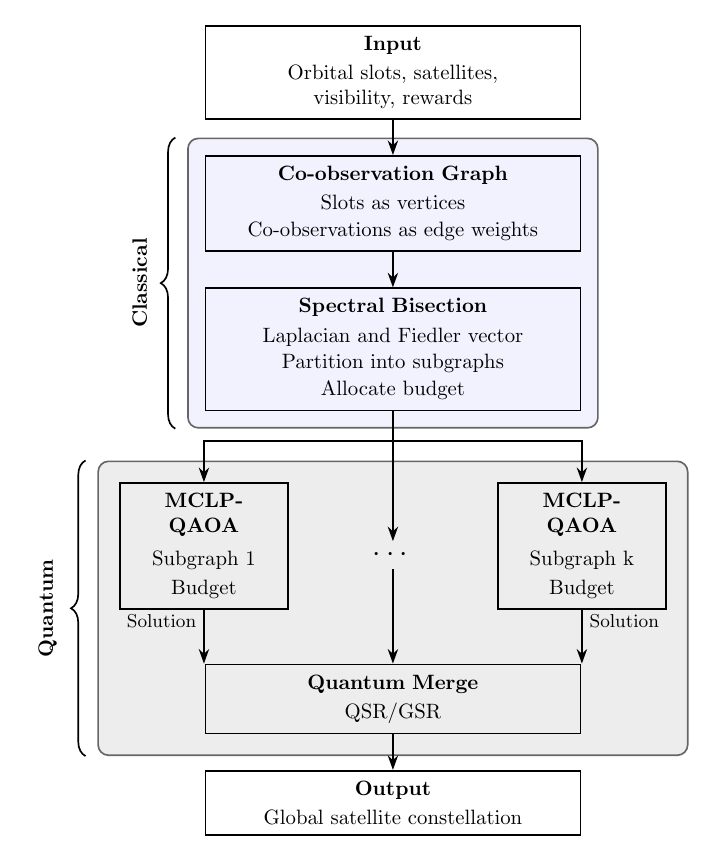}
\caption{Decomposition-Based QAOA Framework Overview.}
\label{fig:framework}
\end{figure}

For MCLP, decomposition must 
preserve coverage semantics. The proposed decomposition-based QAOA framework uses a divide-and-conquer-based strategy. A co-observation graph is constructed
where edge weights encode observational relationships between satellite slots. Decomposing the co-observation graph aligns with the MCLP problem structure and ensures subproblems maintain coverage feasibility. This problem-aware decomposition strategy is specific to coverage-type optimization problems. Fig.~\ref{fig:framework} presents the block diagram of decomposition-based QAOA framework.

The framework comprises four major components. First, the MCLP instance is reformulated as a weighted co-observation graph where vertices represent orbital slots and edge weights encode co-observation patterns. Second, the spectral bisection \cite{fiedler_algebraic_1973, von_luxburg_tutorial_2007, spielman_spectral_1996} partitions the graph into balanced subgraphs that satisfy quantum hardware constraints. Third, each subgraph is solved independently using QAOA with QUBO Hamiltonian formulation \cite{lucas_ising_2014, hadfield_quantum_2019}, producing partial constellations as solutions. Fourth, quantum merging integrates these partial solutions into a globally consistent solution. The following subsections detail each component of the framework.

\subsection{Co-Observation Graph Construction}
\label{sec:graph_construction}
MCLP is mapped into an undirected graph $G$ to enable decomposition, where structural interaction between orbital slots is explicitly modeled. The graph representation transforms the satellite constellation selection problems into a graph partitioning framework, enabling the application of classical spectral methods \cite{spielman_spectral_1996, von_luxburg_tutorial_2007}, within the divide-and-conquer paradigm \cite{li_large-scale_2023}.

\subsubsection{Graph Definition}
The graph $G = (V, W)$ is defined as follows:

\textbf{Vertex Set:} 
Each vertex $v_j \in V$ represents an orbital slot $j \in \{1, 2, \ldots, J\}$ from the MCLP formulation. The vertex set contains $|V| = J$ vertices that represent all candidate orbital slots for satellite placement.

\textbf{Weight Matrix:} 
The symmetric weight matrix $W \in \mathbb{R}^{J \times J}$ encodes the co-observation relationships between orbital slots, with $w_{ij} = w_{ji}$. An edge exists between vertices $i$ and $j$ when $w_{ij} > 0$, indicating that both slots can observe the target at the same time.

\subsubsection{Edge Weight Computation}
Edge weights $w_{ij}$ between orbital slots $i$ and $j$ quantify co-observation frequency, defined as the total number of target-time pairs $(t, p)$ for which both slots have simultaneous visibility:

\begin{equation}
w_{ij} = \sum_{t=1}^{T} \sum_{p=1}^{P} \Phi_{tip} \cdot \Phi_{tjp}
\label{eq:edge_weight}
\end{equation}

\noindent
where the product $\Phi_{tip} \cdot \Phi_{tjp}$ equals 1 if both slots $i$ and $j$ co-observed target $p$ at same time $t$, and 0 otherwise.

Large edge weights $w_{ij}$ indicate strong coverage correlation between slots $i$ and $j$, meaning these slots observe the target at similar time steps. To preserve solution quality during decomposition, orbital slots with high co-observation weights are assigned to the same subgraph, while slots with low co-observation weights are separated across partition boundaries. 

\subsection{Graph Decomposition via Spectral Bisection}
\label{sec:spectral_bisection}
To recursively partition the co-observation graph $G$ into balanced subgraphs, spectral bisection is employed to minimize edge weights across partition boundaries \cite{spielman_spectral_1996,fiedler_algebraic_1973,von_luxburg_tutorial_2007}.

\subsubsection{Objective: Discrete Balanced Min-Cut}

The objective of each bisection step is to partition the vertex set $V$ into two disjoint subsets $V_{m}$ and $V_{n}$ such that $V_{m} \cup V_{n} = V$ and $V_{m} \cap V_{n} = \emptyset$, while minimizing the total weight of edges crossing the partition boundary. Let $E = \{(i,j) : w_{ij} > 0\}$ denote the edge set. The discrete balanced min-cut problem seeks to minimize:
\begin{equation}
\text{Cut}(V_{m}, V_{n}) = \sum_{\substack{i \in V_{m}, j \in V_{n} \\ (i,j) \in E}} w_{ij}
\label{eq:min_cut_discrete}
\end{equation}
subject to the balance constraint $|V_{m}| \approx |V_{n}| \approx J/2$.

This discrete combinatorial optimization problem is NP-hard and computationally intractable for large graphs, requiring the evaluation of exponentially many possible partitions. Spectral bisection provides an efficient approximation through continuous relaxation, as detailed in the following subsections.

\subsubsection{Continuous Relaxation}
Spectral bisection addresses computational intractability 
through continuous relaxation using the graph Laplacian matrix $L \in\mathbb{R}^{J \times J}$, defined as
\begin{equation}
L = D - W
\label{eq:laplacian}
\end{equation}
where $D \in \mathbb{R}^{J \times J}$ represents the degree matrix with the diagonal entries $D_{ii} = \sum_{j=1}^{J} w_{ij}$, and the off-diagonal entries $D_{ij} = 0$ for $i \neq j$.

The relaxed optimization problem seeks the continuous vector $f \in \mathbb{R}^J$ that minimizes:
\begin{equation}
\min_{f \in \mathbb{R}^J} \quad f^T L f \quad \text{subject to} \quad f \perp \mathbf{1}, \; \|f\|_2 = 1
\label{eq:relaxed_mincut}
\end{equation}
where $\mathbf{1}$ denotes the all-ones vector. The orthogonality constraint $f \perp \mathbf{1}$ enforces $\sum_{i=1}^{J} f_i = 0$, ensuring the partition will be balanced. The normalization constraint $\|f\|_2 = 1$ prevents the trivial solution, $f = \mathbf{0}$.

The solution to \eqref{eq:relaxed_mincut} is the eigenvector corresponding to the second smallest eigenvalue of the graph Laplacian $L$, known as the Fiedler vector \cite{fiedler_algebraic_1973, von_luxburg_tutorial_2007}.

\subsubsection{Partition Construction via Median Thresholding}

Let $\mathbf{f}_2 \in \mathbb{R}^J$ the Fiedler vector be obtained by solving the eigenvalue problem $L \mathbf{f}_2 = \lambda_2 \mathbf{f}_2$, where $\lambda_2$ the second smallest eigenvalue of $L$. The Fiedler vector contains continuous real-valued components $f_2(i)$ for each vertex $i \in V$. 

Each spectral bisection step partitions a graph into two subgraphs by thresholding the Fiedler vector at its median value:
\begin{equation}
V_{m} = \{i \in V \mid f_2(i) \leq \text{median}(\mathbf{f}_2)\}, \quad V_{n} = V \setminus V_{m}
\label{eq:partition}
\end{equation}
This produces a balanced bisection with minimal edge-cut weights \cite{spielman_spectral_1996,von_luxburg_tutorial_2007}.

\subsubsection{Recursive Decomposition into k Subgraphs}

To generate $k$ balanced subgraphs satisfying quantum hardware constraints, spectral bisection is applied recursively. Starting from the original graph $G$, each bisection step creates two child subgraphs. The process continues until all subgraphs satisfy the size constraint $|V_\ell| \leq n_{\max}$ for $\ell = 1, \ldots, k$, where $\ell$ indexes the resulting subgraphs and $n_{\max}$ is the maximum number of qubits available for QAOA subproblem optimization.

The recursive partitioning produces $k$ subgraphs $\{G_1, \ldots, G_k\}$ with vertex sets $\{V_1, \ldots, V_k\}$ such that:
\begin{equation}
\bigcup_{\ell=1}^{k} V_\ell = V, \quad V_{m} \cap V_{n} = S_{mn} \; \text{for adjacent subgraphs}
\label{eq:recursive_partition}
\end{equation}
where $S_{mn}$ denotes separator vertices shared between adjacent subgraphs $G_{m}$ and $G_{n}$ in the decomposition hierarchy. The satellite budget $N$ is allocated proportionally across subgraphs as $\{b_1, \ldots, b_k\}$ where $\sum_{\ell=1}^k b_\ell = N$, with each allocation satisfying $b_\ell \leq n_{\max}$ to ensure compatibility with NISQ hardware constraints.

\subsection{Satellite Allocation Strategy}
\label{sec:satellite_strategy}
The total number of satellites, $N$, must be partitioned among subgraphs to enable independent optimization of subproblems. Each subgraph $G_\ell$ is assigned a budget $b_\ell$ representing the number of satellites to be selected from that subgraph. The allocation strategy ensures that budgets are distributed proportionally to subgraph sizes while maintaining the global cardinality constraint $\sum_{\ell=1}^k b_\ell = N$.

\subsubsection{Proportional Allocation}

For each bisection step partitioning a graph with vertex set $V$ and budget $b$ into two subgraphs with vertex sets $V_{m} \cup S$ and $V_{n} \cup S$, where $S$ denotes the separator vertex set shared between subgraphs, the allocated budgets $b_{m}$ and $b_{n}$ are computed as:
\begin{equation}
b_{m} = \left\lfloor b \cdot \frac{|V_{m} \cup S|}{|V|} \right\rfloor, \quad b_{n} = \left\lfloor b \cdot \frac{|V_{n} \cup S|}{|V|} \right\rfloor
\label{eq:budget_allocation}
\end{equation}
where $\lfloor \cdot \rfloor$ denotes the floor function. This proportional allocation reflects each subgraph's relative coverage capacity, ensuring that larger subgraphs receive correspondingly larger budgets.

\subsubsection{Residual Budget Distribution}

Due to rounding in \eqref{eq:budget_allocation}, the sum $b_{m} + b_{n}$ may be less than the parent budget $b$. The residual budget is defined as:
\begin{equation}
r = b - (b_{m} + b_{n})
\label{eq:residual}
\end{equation}
When $r > 0$, the residual satellites are allocated to the larger subgraph to minimize relative imbalance:
\begin{equation}
\begin{cases}
b_{m} \leftarrow b_{m} + r & \text{if } |V_{m} \cup S| \geq |V_{n} \cup S| \\
b_{n} \leftarrow b_{n} + r & \text{if } |V_{m} \cup S| < |V_{n} \cup S|
\end{cases}
\label{eq:residual_allocation}
\end{equation}
This ensures the budget constraint $b_{m} + b_{n} = b$ is satisfied at each decomposition level.

\subsubsection{Recursive Budget Allocation}

For recursive bisection, budget allocation is applied hierarchically at each decomposition level. When a subgraph $G_\ell$ with an allocated budget $b_\ell$ is further partitioned into two child subgraphs $G_{m}$ and $G_{n}$, the budget $b_\ell$ is redistributed proportionally following \eqref{eq:budget_allocation}--\eqref{eq:residual_allocation}. This recursive process continues until all subgraphs satisfy the maximum qubit constraint $|V_\ell| \leq q_{\max}$, yielding $k$ subgraphs with budgets $\{b_1, \ldots, b_k\}$ where $\sum_{\ell=1}^k b_\ell = N$.

The proportional allocation strategy preserves the global optimization objective by distributing satellite resources according to each subgraph's coverage potential, enabling independent subproblem optimization while maintaining consistency with the original MCLP formulation.

\subsection{MCLP-QAOA Subproblem Solver}
\label{sec:MCLP-QAOA}
Once the co-observation graph has been recursively partitioned into subgraphs satisfying the qubit constraint, each subgraph is solved independently using QAOA with QUBO Hamiltonian formulation \cite{lucas_ising_2014,hadfield_quantum_2019}. The subproblem solver encodes the coverage-maximization objective and the cardinality constraint as a quantum optimization problem.

\subsubsection{Classical-to-Quantum Mapping}

The classical MCLP formulation (\ref{eq:objective})--(\ref{eq:coverage_condition}) must be reformulated as a quantum optimization problem suitable for QAOA implementation. This requires three key transformations:

\textbf{Binary Variable Encoding:} Each classical binary decision variable $x_j \in \{0,1\}$ is mapped to the eigenvalue of a Pauli-Z operator $Z_j$ through the relation:
\begin{equation}
x_j = \frac{1 - Z_j}{2}
\end{equation}
where $Z_j |0\rangle = |0\rangle$ corresponds to $x_j = 1$ and $Z_j |1\rangle = -|1\rangle$ corresponds to $x_j = 0$.

\textbf{Coverage Condition Translation:} The classical coverage constraint (\ref{eq:coverage_condition}) states that the target $p$ is covered at time $t$ if at least one satellite with visibility is selected:
\begin{equation}
\sum_{j \in C(t,p)} x_j \geq 1 \quad \Rightarrow \quad y_{tp} = 1
\end{equation}
where $C(t,p) = \{j \in V_\ell \mid \Phi_{tjp} = 1\}$ denotes orbital slots with visibility. In quantum formulation, this translates to requiring that at least one qubit in $C(t,p)$ is in state $|1\rangle$ (selected). The probability that all slots are unselected is:
\begin{equation}
P(\text{no coverage}) = \prod_{j \in C(t,p)} P(x_j = 0) = \prod_{j \in C(t,p)} \frac{1 - Z_j}{2}
\end{equation}
Therefore, the coverage indicator becomes:
\begin{equation}
y_{tp} = 1 - \prod_{j \in C(t,p)} \frac{1 - Z_j}{2}
\end{equation}

\textbf{Objective Function as Energy Minimization:} QAOA minimizes energy, while MCLP maximizes coverage. The objective function (\ref{eq:objective}) is transformed as:
\begin{equation}
\max F = \sum_{t,p} \pi_{tp} \cdot y_{tp} \quad \Rightarrow \quad \min \left( -\sum_{t,p} \pi_{tp} \cdot y_{tp} \right)
\end{equation}

Combining these transformations yields the coverage reward term in the Hamiltonian.

\subsubsection{MCLP Hamiltonian Formulation}
For a subgraph $G_\ell = (V_\ell, W_\ell)$ with $n = |V_\ell|$ vertices and an allocated budget $b_\ell$, the MCLP problem Hamiltonian is encoded as an Ising Hamiltonian acting on $n$ qubits. Each qubit $j \in V_\ell$ corresponds to an orbital slot in the subgraph, where computational basis states $|0\rangle$ and $|1\rangle$ represent the binary decision variable $x_j \in \{0, 1\}$ through the Pauli-Z operator eigenvalue relation $Z_j |0\rangle = |0\rangle$ and ,$Z_j |1\rangle = -|1\rangle$ with the mapping $(1 - Z_j)/2 = x_j$.

The MCLP Hamiltonian combines coverage reward terms with a cardinality penalty:

\begin{equation}
\begin{split}
H_{\text{MCLP}} = &\sum_{t \in T_\ell} \sum_{p=1}^{P} \pi_{tp} \left( 1 - \prod_{j \in C(t,p)} \frac{1 - Z_j}{2} \right) \\
&+ \beta \left( \sum_{j \in V_\ell} \frac{1 - Z_j}{2} - b_\ell \right)^2
\end{split}
\label{eq:mclp_hamiltonian}
\end{equation}

where $C(t, p) = \{j \in V_\ell \mid \Phi_{tjp} = 1\}$ denotes the set of orbital slots in the subgraph $\ell$ that can observe the target $p$ at time $t$, $T_\ell$  represents the set of timesteps for which the subgraph $\ell$ has coverage capability, and $\beta > 0$ is the penalty coefficient enforcing the budget constraint $\sum_{j \in V_\ell} x_j = b_\ell$.

The first term encodes coverage rewards when the target $p$ is covered at time $t$, the product term approaches zero, minimizing the Hamiltonian. The second term penalizes deviations from the budget, ensuring exactly $b_\ell$ satellites are selected.

\subsubsection{QUBO Reformulation}

For a subgraph $G_\ell$ with $n = |V_\ell|$ vertices, QUBO encoding requires exactly $n$ qubits, 
with the constraint $|V_\ell| \leq q_{\max}$. Direct implementation of the product term in \eqref{eq:mclp_hamiltonian} requires multi-body interaction terms that scale exponentially with the coverage set size $|C(t,p)|$. To enable efficient circuit implementation, the Hamiltonian is reformulated as a QUBO problem with pairwise interactions.

\textbf{Product Term Linearization:} The product term 
\begin{equation}
\prod_{j \in C(t,p)} \frac{1-Z_j}{2}
\end{equation}
in the coverage reward is linearized for practical implementation:
\begin{equation}
1 - \prod_{j \in C(t,p)} \frac{1 - Z_j}{2} \approx \sum_{j \in C(t,p)} \frac{1 - Z_j}{2 |C(t,p)|}
\label{eq:coverage_linearization}
\end{equation}
This approximation distributes the coverage reward equally among all slots that can observe the target, avoiding exponential scaling while preventing over-counting of redundant coverage. To quantify the accuracy of the proposed linearization, the approximation error is bounded by:

\begin{equation}
\left| 1 - \prod_{j \in C(t,p)} \frac{1-Z_j}{2}
- \sum_{j \in C(t,p)}
\frac{1 - Z_j}{2|C(t,p)|} \right|
\leq \frac{1}{2} - \left(\frac{1}{2}\right)^{|C(t,p)|}
\label{eq:error_bound}
\end{equation}

The bound guarantees that the linearization introduces a controlled approximation error that depends solely on the number of candidate observing orbital slots.

\textbf{Cardinality Penalty Expansion:} The quadratic cardinality constraint is expanded by substituting $x_i = (1-Z_i)/2$:
\begin{equation}
\begin{split}
\left(\sum_i x_i - b_\ell\right)^2 &= \sum_i x_i^2 + 2\sum_{i<j} x_i x_j \\
&\quad - 2b_\ell \sum_i x_i + b_\ell^2
\end{split}
\end{equation}
Since $x_i \in \{0,1\}$ implies $x_i^2 = x_i$, this simplifies to linear diagonal terms $\sum_i x_i$ and quadratic off-diagonal terms $2\sum_{i<j} x_i x_j$.

\textbf{QUBO Matrix Construction:} Combining the linearized coverage terms and expanded cardinality penalty, the QUBO matrix $Q \in \mathbb{R}^{n \times n}$ for subgraph $G_\ell$ with $n = |V_\ell|$ vertices is defined as:
\begin{equation}
Q_{ij} = \begin{cases}
-\displaystyle\sum_{t \in T_\ell} \sum_{\substack{p: i \in C(t,p)}} \frac{\pi_{tp}}{|C(t,p)|} \\
\quad + \beta(1 - 2b_\ell) & \text{if } i = j \\[10pt]
\beta & \text{if } i \neq j
\end{cases}
\label{eq:qubo_matrix}
\end{equation}
where $i, j \in \{1, \ldots, n\}$ index the vertices within subgraph $\ell$. The cardinality penalty coefficient is set to $\beta = 50$ to balance constraint satisfaction with solution quality. This value is selected via grid search over $\beta \in [10, 50, 100, 500]$ to minimize constraint violations while maintaining coverage performance across all test instances.

The diagonal elements $Q_{ii}$ aggregate coverage contributions from all target-time pairs where slot $i$ provides visibility, weighted inversely by coverage set size to avoid over-counting shared coverage. The linear term $\beta(1-2b_\ell)$ arises from the $-2b_\ell \sum_i x_i$ term in the cardinality penalty expansion. The off-diagonal elements $Q_{ij} = \beta$ implement the quadratic cardinality penalty through the cross terms $2\sum_{i<j} x_i x_j$, which when combined with the diagonal linear terms, reproduces the constraint $(\sum_i x_i - b_\ell)^2$ after expansion.

The QUBO formulation minimizes the following:
\begin{equation}
\min_{x \in \{0,1\}^n} \quad x^T Q x
\label{eq:qubo_objective}
\end{equation}

This formulation enables standard QAOA circuit compilation with Pauli-Z interaction terms $Z_i Z_j$ corresponding to QUBO matrix elements $Q_{ij}$.

\subsubsection{QAOA circuit and Optimization}

The QAOA circuit prepares a parametrized quantum state through alternating applications of the problem Hamiltonian and mixer operators over $p$ layers:

\begin{equation}
|\psi(\gamma, \beta)\rangle = \prod_{r=1}^{p} \left(e^{-i\beta_r \sum_j X_j} \cdot e^{-i\gamma_r H_{\text{MCLP}}}\right) |+\rangle^{\otimes n}
\label{eq:qaoa_circuit}
\end{equation}

\noindent
where $|+\rangle^{\otimes n}$ denotes the uniform superposition initial state, $X_j$ are Pauli-X operators providing quantum mixing. Variational parameters are initialized as $\gamma_r \sim U[0, \pi]$ and $\beta_r \sim U[0, \pi/2]$ for all $r = 1, \ldots, p$. Classical optimization employs the constrained optimization by linear approximation (COBYLA) algorithm with maximum 100 iterations for base MCLP-QAOA cases. Each circuit execution uses 10,000 measurement shots to ensure statistical convergence of the probability 
distribution. Optimization convergence is assessed when the expected energy change falls below $10^{-4}$ across consecutive iterations or the maximum iteration count is reached.

\begin{figure}[h]
\centering 
\Large
\includegraphics[width=\columnwidth]{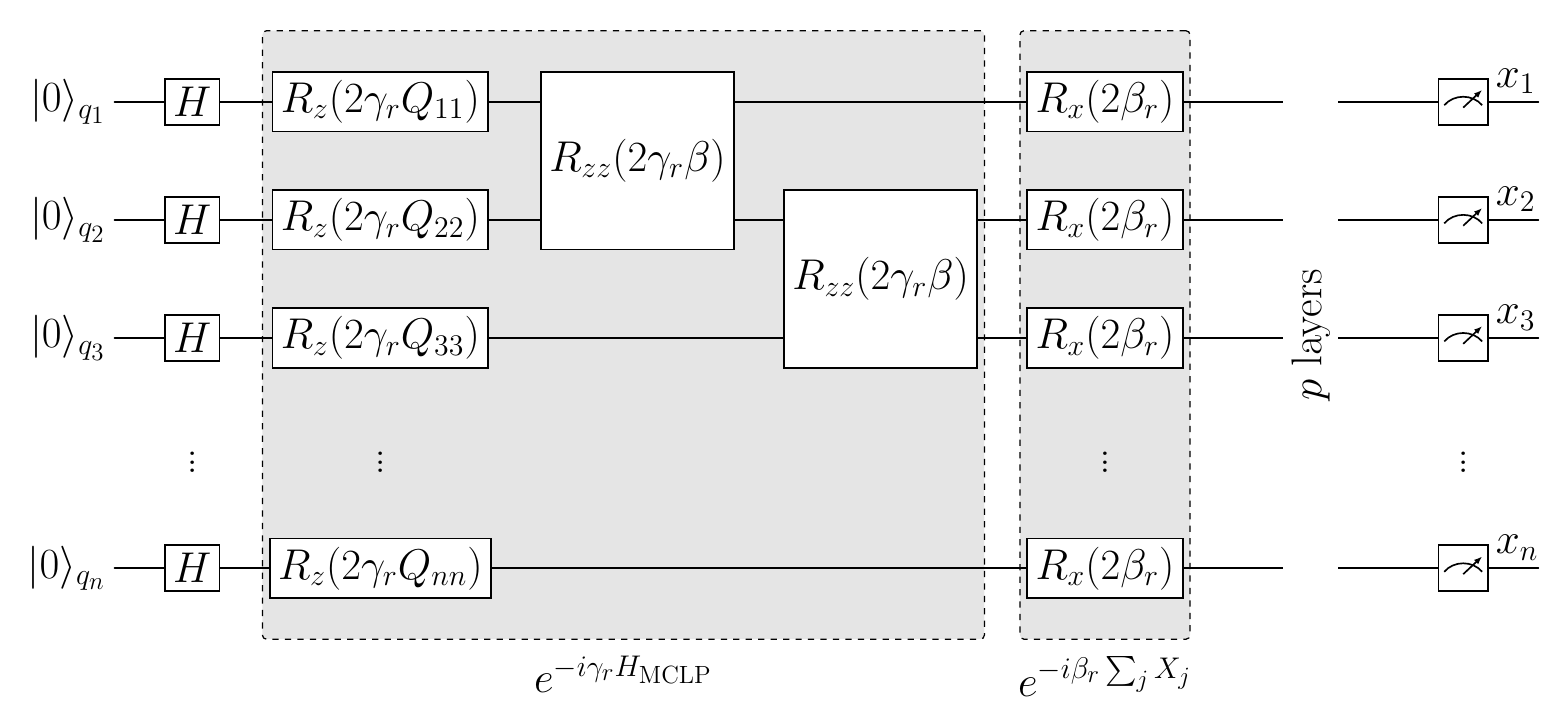}
\caption{\footnotesize MCLP-QAOA circuit for subgraph $G_\ell$ with $p$ problem and mixer layers using optimized $(\gamma_r, \beta_r)$ parameters.}
\label{fig:mclp-qaoa-circuit}
\end{figure}

Fig.~\ref{fig:mclp-qaoa-circuit} illustrates an MCLP-QAOA circuit for a subgraph $G_\ell$
with $n$ qubits and $p$ layers. Each qubit $q_j$ represents one
orbital slot and $x_j\in\{0,1\}$ indicates whether it is selected.
$H$ Gates prepare the uniform superposition $|{+}\rangle^{\otimes n}$.
Each layer $r$ applies single-qubit $R_z(2\gamma_r Q_{jj})$
gates encoding the coverage reward and cardinality penalty (diagonal
QUBO entries $Q_{jj}$), two-qubit $R_{zz}(2\gamma_r\beta)$
gates on all $\binom{n}{2}$ qubit pairs enforcing the budget
constraint ($Q_{ij}=\beta$, $i\neq j$); and, mixer
$R_x(2\beta_r)$ on each qubit independently. Dashed boxes group
the problem unitary $e^{-i\gamma_r H_{\mathrm{MCLP}}}$ and
mixer $e^{-i\beta_r\sum_j X_j}$ within each layer. The circuit structure repeats for $p$ layers, with different parameters $(\gamma_r, \beta_r)$. The feasible bitstring $\mathbf{x}^*$ with the highest probability satisfying it $\sum_j x_j=b_\ell$ is selected as a subgraph constellation $\mathcal{C}_\ell$.

The optimization objective minimizes the expected energy:
\begin{equation}
\min_{\gamma, \beta} \quad \langle \psi(\gamma, \beta) | H_{\text{MCLP}} | \psi(\gamma, \beta) \rangle
\label{eq:qaoa_optimization}
\end{equation}
After classical optimization converges to optimal parameters $(\gamma^*, \beta^*)$, the quantum state $|\psi^*\rangle = |\psi(\gamma^*, \beta^*)\rangle$ is measured in a computational basis to obtain a probability distribution over bit strings:
\begin{equation}
P(x) = |\langle x | \psi^* \rangle|^2
\label{eq:qaoa_distribution}
\end{equation}

\subsubsection{Solution Extraction}
The subgraph constellation $\mathcal{C}_\ell$ is selected from the measurement distribution by identifying the feasible bit string with the highest probability that satisfies the budget constraint. Feasible solutions satisfy $\sum_{j=1}^n x_j = b_\ell$, ensuring exactly $b_\ell$ satellites are allocated to subgraph $\ell$. Among feasible bitstrings, the solution maximizing the MCLP coverage objective \eqref{eq:objective} is selected as the subgraph constellation.

This QAOA-based subproblem solver scales to subgraphs with $q \leq q_{\text{max}}$ qubits, enabling quantum optimization of moderate-sized MCLP instances.

\subsection{Quantum Merge}
\label{sec:quantum_merge}
The partial constellation solutions obtained from recursive decomposition and independent QAOA optimization of subgraphs need to be merged for a globally consistent satellite configuration. The merge operation addresses the challenge of reconciling potentially conflicting separator node assignments, where the orbital slot may be selected by one subgraph and excluded by another. 
Two quantum merging strategies are presented: QSR using replicated separator qubits with post-selection and GSR using coverage-weighted optimization on the separator subgraph.

\subsubsection{Quantum State Reconstruction (QSR)}

QSR enforces agreement between subproblem solutions by representing each separator slot with two qubits, one for each adjacent subgraph, and applying penalty terms to encourage identical assignments. The method operates on $2|S|$ qubits, where $|S|$ denotes the separator size.

The QSR Hamiltonian combines agreement penalties with co-observation coupling:
\begin{equation}
\begin{split}
H_{\text{QSR}} &= \sum_{i \in S} \lambda \cdot Z_{i1} Z_{i2} \\
&\quad + \sum_{i<j \in S} w_{ij} \left(Z_{i1} Z_{j1} + Z_{i2} Z_{j2}\right)
\end{split}
\label{eq:qsr_hamiltonian}
\end{equation}
where $Z_{i1}$ and $Z_{i2}$ represent Pauli-Z operators on the two qubit copies for separator slot $i$, $\lambda > 0$ is the agreement penalty coefficient, and $w_{ij}$ denotes the co-observation weight between slots $i$ and $j$ from the global weight matrix $W$.

The first term penalizes disagreement between qubit copies, achieving 
minimum energy $Z_{i1} = Z_{i2}$ for all separator slots. The second term preserves co-observation structure by coupling slots with strong interaction weights within each subgraph representation.

For QUBO formulation, each separator slot $i$ uses two qubits with indices $q_1 = i$ and $q_2 = i + |S|$. For a separator with $|S|$ vertices, the QUBO matrix requires 
$2|S|$ qubits. The QUBO matrix $Q \in \mathbb{R}^{2|S| \times 2|S|}$ incorporates agreement penalties:

\begin{align}
Q_{q_1, q_1} &\leftarrow Q_{q_1, q_1} - \lambda \nonumber \\
Q_{q_2, q_2} &\leftarrow Q_{q_2, q_2} - \lambda \\
Q_{q_1, q_2} &\leftarrow Q_{q_1, q_2} + 2\lambda \nonumber
\end{align}

Co-observation terms are added symmetrically to both qubit copies for each pair $(i,j)$ where $i < j \in S$:
\begin{align}
Q_{i,i} &\leftarrow Q_{i,i} - w_{ij} \nonumber \\
Q_{j,j} &\leftarrow Q_{j,j} - w_{ij} \\
Q_{i,j} &\leftarrow Q_{i,j} + 2w_{ij} \nonumber
\end{align}
applied to both the first $|S|$ qubits (copy 1) and qubits $|S|+1$ through $2|S|$ (copy 2).
The agreement penalty coefficient is set to $\lambda = 100$ to enforce strong agreement between qubit copies. Post-selection retains only measurement outcomes satisfying $s^{(1)} = s^{(2)}$, ensuring separator consistency across adjacent subgraphs. Quantum merge optimization uses COBYLA with 500 maximum iterations and 5,000 measurement shots.

The QAOA circuit prepares the quantum state:
\begin{equation}
|\Phi(\gamma', \beta')\rangle = \prod_{r=1}^{p_{\text{merge}}} \left(e^{-i\beta'_r \sum_q X_q} \cdot e^{-i\gamma'_r H_{\text{QSR}}}\right) |+\rangle^{\otimes 2|S|}
\label{eq:qsr_circuit}
\end{equation}
where $r$ indexes the merge layers with parameters $\{\gamma'_r, \beta'_r\}_{r=1}^{p_{\text{merge}}}$.

\begin{figure}[h]
\centering
\includegraphics[width=\columnwidth]{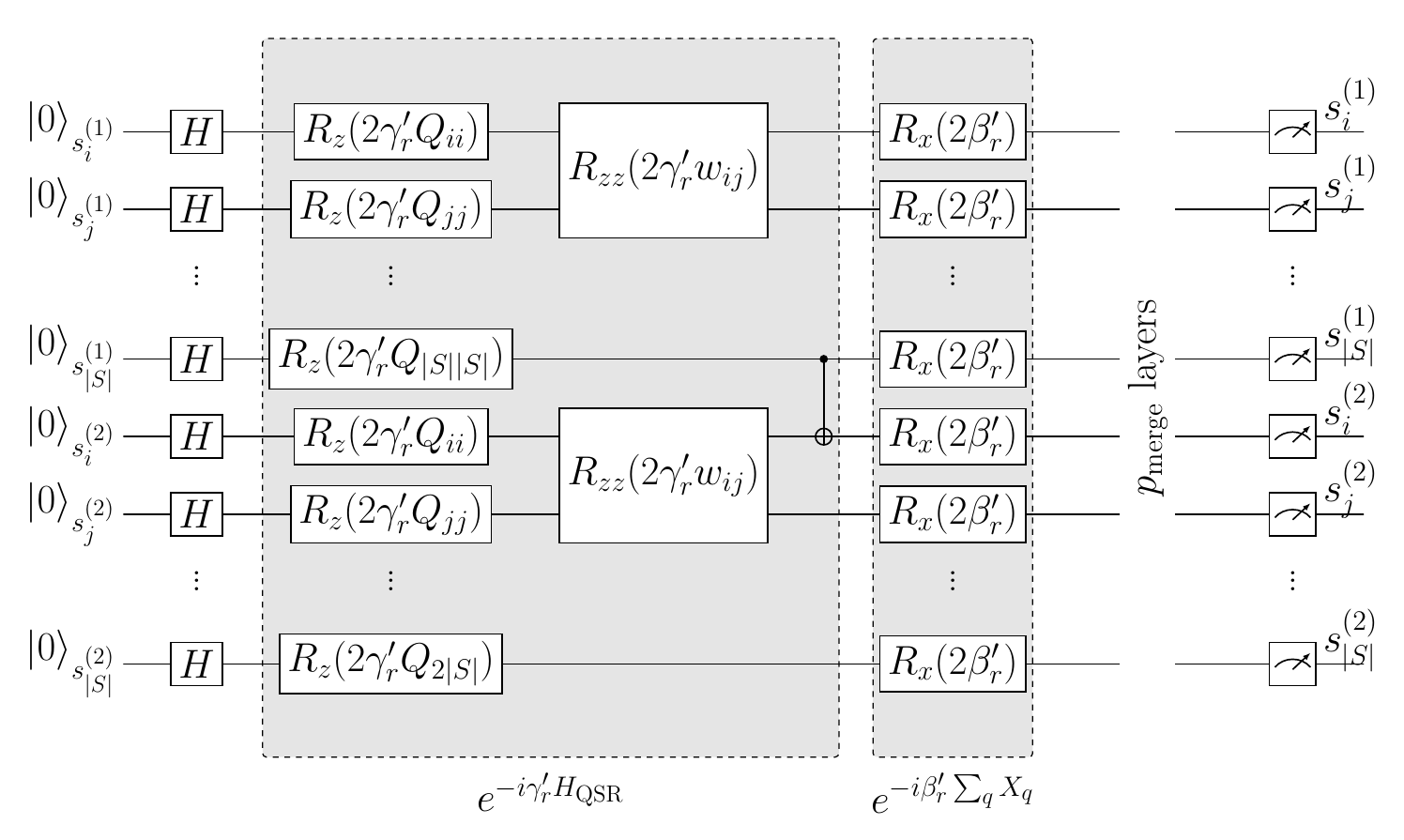}
\caption{\footnotesize QSR circuit with two separator copies applying $p_{\text{merge}}$ problem and mixer layers using cross-copy CNOTs and $R_{zz}$ co-observation coupling gates.}
\label{fig:qsr-circuit}
\end{figure}

Fig.~\ref{fig:qsr-circuit} shows a QSR circuit with $2|S|$ qubits and $p_{\text{merge}}$ layers. $s_i^{(1)}$ and $s_i^{(2)}$ are two copies of separator qubits representing $S$ adjacent subgraphs. Each layer $p_{\text{merge}}$ applies the problem unitary $e^{-i\gamma'_r H_{\text{QSR}}}$ and mixer unitary $e^{-i\beta'_r\sum_q X_q}$. The first term $\sum_{i\in S}\lambda \cdot Z_{i1}Z_{i2}$ $H_{\text{QSR}}$ enforces agreement between copies via cross-copy CNOT gates, while the second term preserves co-observation structure via $R_{zz}(2\gamma'_r w_{ij})$ gates.

After classical optimization, measurement outcomes are post-selected to retain only states where both qubit copies agree: $s^{(1)} = s^{(2)}$, where $s^{(1)}$ and $s^{(2)}$ represent bit strings from the two copies. The post-selected probability distribution is:
\begin{equation}
P(s) = \frac{|\langle s, s | \Phi^* \rangle|^2}{\sum_{s'} |\langle s', s' | \Phi^* \rangle|^2}
\label{eq:qsr_postselection}
\end{equation}

The separator configuration $s^*_{mn}$ with highest post-selected probability is selected. The merged constellation is constructed as:
\begin{equation}
\mathcal{C}_{mn} = (\mathcal{C}_{m} \setminus S_{mn}) \cup (\mathcal{C}_{n} \setminus S_{mn}) \cup s^*_{mn}
\label{eq:qsr_merge}
\end{equation}
where $\mathcal{C}_{m}$ and $\mathcal{C}_{n}$ represent partial constellations from adjacent subgraphs.

The $2|S_{mn}|$ qubit requirement limits QSR scalability for large separator sets. For practical NISQ implementations, separator sizes must satisfy $2|S_{mn}| \leq q_{\max}$ to remain within available qubit capacity.

\subsubsection{Graph State Reconstruction (GSR)}

GSR addresses QSR's qubit overhead by directly optimizing separator slot selection using a single qubit per separator vertex.The GSR Hamiltonian operates on $|S|$ qubits:

\begin{equation}
\begin{split}
H_{\text{GSR}} &= -\sum_{i \in S} \sigma_i \cdot Z_i + \sum_{i<j \in S} w_{ij} \cdot Z_i Z_j \\
&\quad + \beta \left(\sum_{j \in S} Z_j - (2b - |S|)\right)^2
\end{split}
\label{eq:gsr_hamiltonian}
\end{equation}
where $\sigma_i = \sum_{t \in T} \sum_{p=1}^{P} \pi_{tp} \cdot \Phi_{tip}$ represents the weighted coverage score for separator slot $i$, aggregating priority-weighted visibility across all target-time pairs. The separator budget $b$ is computed as:
\begin{equation}
b_{mn} = N - |\mathcal{C}_{m} \setminus S_{mn}| - |\mathcal{C}_{n} \setminus S_{mn}|
\label{eq:separator_budget}
\end{equation}
ensuring that the merged constellation satisfies the budget constraint for the separator between subgraphs $m$ and $n$.

The first term in \eqref{eq:gsr_hamiltonian} biases selection toward separator slots with high coverage potential. Slots with large $\sigma_i$ values contribute greater negative energy when selected ($Z_i = -1$), favoring configurations that maximize coverage rewards. The second term preserves co-observation structure by rewarding simultaneous selection of strongly coupled slot pairs. The third term enforces the cardinality constraint, penalizing deviations from the separator budget $b$.

The QUBO reformulation yields diagonal and off-diagonal matrix elements:
\begin{align}
Q_{ii} &= -\frac{\sigma_i}{\alpha} + \beta(1 - 2b) \label{eq:gsr_qubo_diag} \\
Q_{ij} &= \frac{w_{ij}}{\alpha} + \beta \quad \text{for } i \neq j \label{eq:gsr_qubo_offdiag}
\end{align}
where the cardinality penalty is $\beta = 50$ and the normalization coefficient is $\alpha = 1$. These values balance coverage bias from $\sigma_i$ with co-observation coupling $w_{ij}$ and cardinality enforcement. GSR quantum 
merge optimization uses COBYLA with 500 maximum iterations and 5,000 
measurement shots, with no post-selection required due to the single-copy qubit encoding. The reason behind increased number of iterations and reduced measurement shots is because merge operations optimize on smaller separator subgraphs with tighter agreement/cardinality constraints, requiring more number of COBYLA iterations for fine-tuning despite reduced problem size, yet fewer or half measurement shots suffice for statistical convergence in the smaller state space (compared to 100 iterations and 10,000 shots for base MCLP-QAOA subproblems).
For a separator with $|S|$ vertices, the QUBO matrix is $Q \in \mathbb{R}^{|S| \times |S|}$, 
requiring $|S|$ qubits for the merge operation, half the qubit count of QSR ($2|S|$ qubits) while achieving comparable or superior solution quality.

The QAOA circuit for GSR prepares the state:
\begin{equation}
|\Psi(\gamma'', \beta'')\rangle = \prod_{r=1}^{p_{\text{merge}}} \left(e^{-i\beta''_r \sum_{j \in S} X_j} \cdot e^{-i\gamma''_r H_{\text{GSR}}}\right) |+\rangle^{\otimes |S|}
\label{eq:gsr_circuit}
\end{equation}
where $r$ indexes the merge layers with parameters $\{\gamma''_r, \beta''_r\}_{r=1}^{p_{\text{merge}}}$.

\begin{figure}[h]
\centering
\includegraphics[width=\columnwidth]{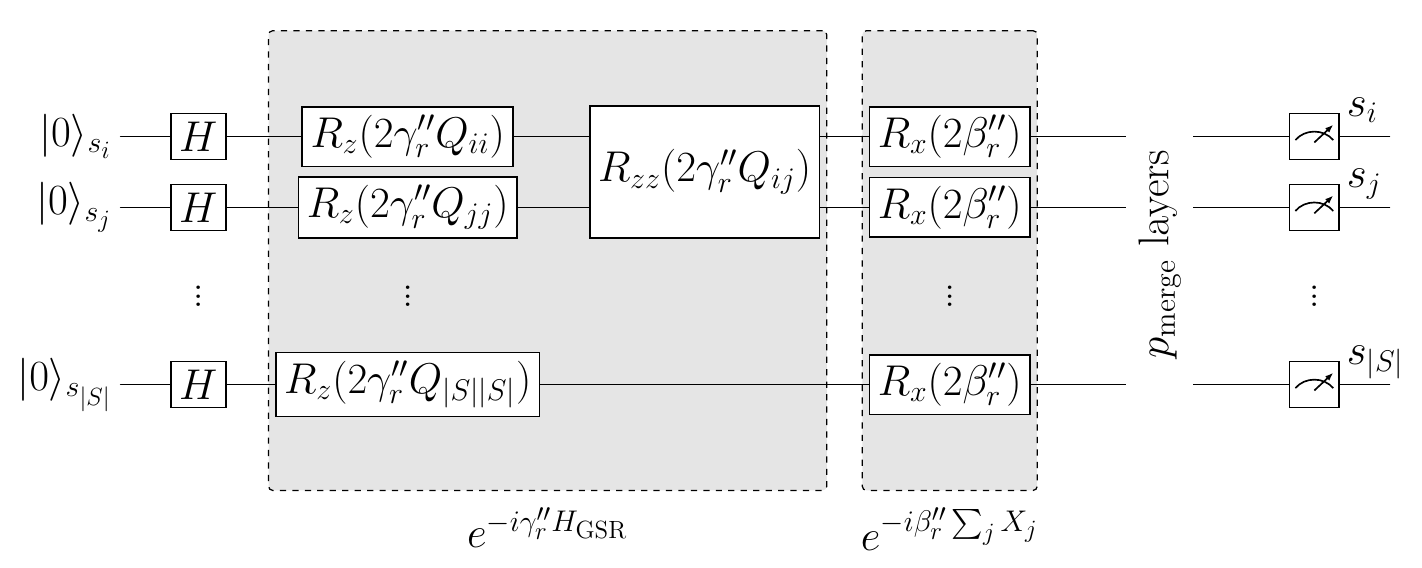}
\caption{\footnotesize GSR circuit applying $p_{\text{merge}}$ problem and mixer layers with coverage bias, co-observation coupling, and cardinality constraints using $R_z$ and $R_{zz}$ gates.}
\label{fig:gsr-circuit}
\end{figure}

Fig.~\ref{fig:gsr-circuit} shows GSR circuit with $|S|$ qubits and $p_{\text{merge}}$ layers. Each qubit $s_j$ represents one separator qubit, requiring half the qubits of QSR with no post-selection. Each layer of $r$ applies the problem unitary $e^{-i\gamma''_r H_{\text{GSR}}}$ and mixer $e^{-i\beta''_r\sum_j X_j}$. Single-qubit $R_z(2\gamma''_r Q_{ii})$ gates encode coverage bias and cardinality penalty, while two-qubit $R_{zz}(2\gamma''_r Q_{ij})$ gates encode co-observation coupling and cardinality constraints. Mixer $R_x(2\beta''_r)$ gates complete each layer.

After variational optimization, the separator configuration is selected by identifying the bit string with correct cardinality and maximum probability:
\begin{equation}
s^* = \operatorname*{argmax}_{s: \sum_{j=1}^{|S|} s_j = b} P(s)
\label{eq:gsr_selection}
\end{equation}
where $P(s) = |\langle s | \Psi^* \rangle|^2$ represents the measurement probability distribution.

The merged constellation is constructed identically to \eqref{eq:qsr_merge}:
\begin{equation}
\mathcal{C}_{mn} = (\mathcal{C}_{m} \setminus S_{mn}) \cup (\mathcal{C}_{n} \setminus S_{mn}) \cup s^*_{mn}
\label{eq:gsr_merge}
\end{equation}
where non-separator vertices from both subgraphs are combined via set union, and separator slots are determined by GSR optimization.

In contrast to QSR, which requires agreement between subproblem selections over the entire separator using $2|S|$ qubits, 
GSR is a problem-aware merge strategy designed specifically for MCLP coverage structure. GSR exploits the satellite 
co-observation graph weights $w_{ij}$ to optimize complementary coverage between 
partial constellations $C_m$ and $C_n$.
The reduced qubit requirement of GSR enables separator sizes up to $|S| \leq q_{\max}$, doubling the maximum separator capacity relative to QSR's $2|S| \leq q_{\max}$ constraint. This scalability advantage permits coarser graph partitioning, reducing recursion depth and overall decomposition overhead in the framework.

\subsection{Algorithm Description}
\label{sec:algorithm_description}
This subsection presents the algorithm that constructs the co-observation graph from visibility data, recursively partitions it into hardware-compatible subgraphs, solves each subgraph independently using QAOA, and merges the partial solutions hierarchically via quantum merge operations on separator vertices. It also provides the computational complexity analysis of the proposed decomposition-based framework.

\subsubsection{Decomposition-Based MCLP-QAOA Algorithm}

\begin{algorithm}[htbp]
\caption{Decomposition-Based MCLP-QAOA}
\label{alg:dc_qaoa}
\begin{algorithmic}[1]
\STATE \textbf{Input:} Visibility matrix $\Phi \in \{0,1\}^{T \times J \times P}$, satellites \\
$N$, max qubits $q_{\max}$
\STATE \textbf{Output:} Merged constellation $\mathcal{C}^*$ with $|\mathcal{C}^*| = N$
\STATE Construct co-observation graph $G = (V, W)$: $w_{ij} \leftarrow \sum_{t,p} \Phi_{tip} \cdot \Phi_{tjp}$ for all $i,j \in V$
\STATE Decompose $G$ into $k$ subgraphs $\{G_1, \ldots, G_k\}$ via recursive spectral bisection
\STATE Allocate budgets $\{b_1, \ldots, b_k\}$ proportionally with $\sum_{\ell=1}^k b_\ell = N$
\STATE \textbf{for} $\ell = 1$ to $k$ \textbf{do}
\STATE \quad Construct QUBO matrix $Q_\ell$ from MCLP Hamiltonian for $G_\ell$
\STATE \quad Run QAOA: optimize $\{\gamma_r, \beta_r\}_{r=1}^p$ to minimize expected energy $\langle H_{\text{MCLP}} \rangle$
\STATE \quad Measure $|\psi^*\rangle$ and obtain optimal solution $\mathcal{C}_\ell$
\STATE \textbf{end for}
\STATE Store partial constellations: $\mathcal{P} \leftarrow \{\mathcal{C}_1, \ldots, \mathcal{C}_k\}$
\STATE \textbf{for} each adjacent subgraph pair $(G_{m}, G_{n})$ in $\{G_1, \ldots, G_k\}$ \textbf{do}
\STATE \quad Extract separator $S_{mn} = V_{m} \cap V_{n}$
\STATE \quad Compute separator budget $b_{mn} = N - |\mathcal{C}_{m} \setminus S_{mn}| - |\mathcal{C}_{n} \setminus S_{mn}|$
\STATE \quad Apply quantum merge (QSR or GSR) on $S_{mn}$ to obtain $s^*_{mn}$ with $|s^*_{mn}| = b_{mn}$
\STATE \quad Construct merged constellation: $\mathcal{C}_{mn} \leftarrow (\mathcal{C}_{m} \setminus S_{mn}) \cup (\mathcal{C}_{n} \setminus S_{mn}) \cup s^*_{mn}$
\STATE \quad Update: $\mathcal{P} \leftarrow (\mathcal{P} \setminus \{\mathcal{C}_{m}, \mathcal{C}_{n}\}) \cup \{\mathcal{C}_{mn}\}$
\STATE \textbf{end for}
\STATE $\mathcal{C}^* \leftarrow$ the single remaining constellation in $\mathcal{P}$
\STATE \textbf{return} $\mathcal{C}^*$
\end{algorithmic}
\end{algorithm}

Algorithm~\ref{alg:dc_qaoa} describes the complete decomposition-based MCLP-QAOA. The algorithm begins by constructing the co-observation graph $G = (V, W)$ where edge weights $w_{ij}$ encode co-observation frequencies between orbital slots. Recursive spectral bisection decomposes $G$ into $k$ subgraphs $\{G_1, \ldots, G_k\}$ satisfying the qubit constraint $|V_\ell| \leq q_{\max}$. The satellite budget $N$ is allocated proportionally across subgraphs as $\{b_1, \ldots, b_k\}$ with $\sum_{\ell=1}^k b_\ell = N$.

Each subgraph $G_\ell$ is solved independently via QAOA: the MCLP Hamiltonian is reformulated as a QUBO matrix $Q_\ell$, variational layer parameters $\{\gamma_r, \beta_r\}_{r=1}^p$ are optimized classically, and the highest-probability feasible bit string satisfying $|\mathcal{C}_\ell| = b_\ell$ is selected as the partial constellation.

The hierarchical merge process addresses conflicting assignments at separator vertices $S_{mn} = V_{m} \cap V_{n}$ between adjacent subgraphs. For each separator, the budget $b_{mn}$ is computed to maintain the global constraint. Quantum merge is applied using either QSR (operating on $2|S_{mn}|$ qubits with post-selection enforcing $s^{(1)} = s^{(2)}$) or GSR (operating on $|S_{mn}|$ qubits without post-selection) to obtain the optimal separator configuration $s^*_{mn}$. The merged constellation is constructed as $\mathcal{C}_{mn}$.

The hierarchical merge process maintains a constellation set $\mathcal{P}$ containing all partial solutions. After each merge operation, the parent constellations $\mathcal{C}_{m}$ and $\mathcal{C}_{n}$ are removed from the set and replaced by their merged result $\mathcal{C}_{mn}$, reducing the number of partial constellations by one. This process continues hierarchically following the decomposition tree structure until all subgraphs have been merged. The final constellation $\mathcal{C}^*$ is the single remaining constellation in the set, satisfying the global cardinality constraint $|\mathcal{C}^*| = N$ by construction.

Although graph decomposition enables scalable optimization, it may introduce approximation errors during the merging stage due to interactions across graph separators $S_{mn}$. Each subgraph $G_{\ell}$ is solved independently using QAOA while satisfying the local cardinality constraint $\sum_{i \in V_{\ell}} x_i = b_{\ell}$. During reconstruction, the separator budget is computed according to (\mbox{\ref{eq:separator_budget}}) and enforced through the proposed QSR or GSR procedure. Consequently, only the separator variables $s_{mn}^{*}$ are re-optimized while the remaining subgraph solutions are preserved. This construction guarantees that the merged constellation satisfies the global cardinality constraint $|C^{*}|=N$, thereby preserving feasibility throughout the reconstruction process. Since optimization is restricted to separator nodes rather than the entire solution space, any approximation introduced during merging remains localized to the separator regions, thereby limiting the propagation of reconstruction errors across successive merging stages.

\subsubsection{Computational Complexity Analysis}

The computational complexity of the proposed framework is governed by the graph decomposition strategy and the subsequent quantum optimization of the resulting subproblems. For an MCLP instance with $n$ decision variables ($n=J$), standard QAOA formulates the optimization problem on the complete graph and therefore requires $O(n)$ qubits. As the problem size increases, the corresponding quantum circuit rapidly exceeds the qubit capacity of current NISQ hardware.

 The proposed decomposition-based framework recursively partitions the co-observation graph into $k$ balanced subgraphs, each containing at most $q_{\max}$ vertices, where $q_{\max}$ denotes the maximum number of qubits available on the target quantum device. Consequently, each QAOA subproblem requires only $O(q_{\max})$ qubits, independent of the total problem size $n$. The number of subgraphs scales approximately as $k=O(n/q_{\max})$.

The overall computational cost consists of three stages: (i) recursive graph decomposition using spectral bisection, (ii) independent QAOA optimization of each subgraph, and (iii) hierarchical merging of partial solutions. Since the size of each subproblem is bounded by $q_{\max}$, the quantum optimization cost of each QAOA execution remains bounded, while the total computational cost grows primarily with the number of subgraphs. Therefore, the proposed framework improves the practical scalability of QAOA by reducing the quantum resource requirement from $O(n)$ qubits to $O(q_{\max})$ qubits per optimization instance, enabling larger MCLP problems to be executed on current quantum hardware.

\section{Experimental Setup}
\label{sec:experimental_setup}
This section describes the experimental setup for evaluating decomposition-based QAOA on satellite constellation design problems. The MCLP instances are defined based on repeat ground-track orbital mechanics, including visibility matrix generation and parameter configurations. Solver configurations are detailed for both classical optimization using Gurobi and quantum approaches using decomposition-based MCLP-QAOA  with QSR and GSR merge strategies. Performance metrics are specified to assess solution quality, computational cost, and scalability as problem size increases.

\subsection{Problem Instances}
\label{sec:problem_instances}
To evaluate our approach on MCLP instances representing satellite constellation design scenarios, the problem setup is based on a repeat ground-track orbit with parameters derived from orbital mechanics. Each instance consists of orbital slots distributed evenly across the orbital plane, ground targets on Earth's surface, and a satellite cardinality constraint. The visibility matrix $\Phi_{tjp} \in \{0,1\}$ is determined by satellite-to-target visibility using a conical sensor model. Priority values $\pi_{tp}$ represent the relative importance of observing the target at each time step. Co-observation graphs $G = (V, W)$ quantify shared coverage between slots. 

The orbital configuration employs a repeat ground-track orbit. The repeat ground-track period $T_r$ is computed using $J_2$ perturbation theory, accounting for nodal regression and argument of perigee drift. The semi-major axis is iteratively determined to satisfy the repeat ground-track condition, wherein the satellite completes $N_p$ orbital periods in $N_d$ nodal days, ensuring that the ground track repeats after each cycle. Time discretization is performed using fixed intervals $\Delta t$ over the complete repeat cycle, yielding $T = \lfloor T_r / \Delta t \rfloor$ discrete time steps. Orbital mechanics and visibility matrix generation are implemented in MATLAB R2023b. Parameters are summarized in Table~\ref{tab:orbital_params}.

\renewcommand{\thetable}{\Roman{table}}

\begin{table}[H]
\centering
\renewcommand{\arraystretch}{1.1}
\setlength{\tabcolsep}{6.5pt}
\caption{Orbital mechanics parameters for visibility matrix generation in MATLAB.}
\begin{tabular}{|l|c|c|}
\hline
\textbf{Parameter} & \textbf{Symbol} & \textbf{Value} \\
\hline
Orbital periods per cycle & $N_p$ & $[5, 8]$ \\
Nodal days per cycle & $N_d$ & 1 \\
Orbital inclination & $i$ & $[60^{\circ}, 90^{\circ}]$ \\
Eccentricity & $e$ & 0 \\
RAAN at epoch & $\Omega_0$ & $50^\circ$  \\
Argument of latitude at epoch & $u_0$ & $0^\circ$ \\
Time step interval & $\Delta t$ & $[60,300]$ s \\
Sensor max view angle & $\theta_{\max}$ & 90° \\
Latitude & $\phi_{\text{ref}}$ & 37.23$^\circ$ N \\
Longitude & $\lambda_{\text{ref}}$ & 80.41$^\circ$ W \\
Earth radius & $R_{\oplus}$ & 6378.137 km \\
Earth gravitational parameter & $\mu_{\oplus}$ & 398600.4415 km³/s² \\
Earth rotation rate & $\omega_{\oplus}$ & $2\pi/86164.1$ rad/s \\
J2 coefficient & $J_2$ & 0.0010826269 \\
\hline
\end{tabular}
\label{tab:orbital_params}
\end{table}

Three key parameters are varied to evaluate performance under different orbital configurations: the number of orbital periods per cycle $N_p \in [5, 8]$, the orbital inclination $i \in [60^{\circ}, 90^{\circ}]$, and the time step interval $\Delta t \in [60, 300]$ seconds. Varying $N_p$ changes both the number of orbital slots and the repeat cycle duration. The inclination affects latitude coverage, with higher inclinations providing improved coverage of polar regions, while $\Delta t$ controls the temporal resolution of the visibility analysis.

The orbit eccentricity $e$ is fixed for circular orbit. The right ascension of the ascending node (RAAN) $\Omega_0$ and the argument of latitude $u_0$ are assigned fixed values to define the initial orbital geometry for simulation consistency. The sensor follows a conical field-of-view model with a maximum view angle $\theta_{\max}$. Earth parameters include the equatorial radius $R_\oplus$, gravitational parameter $\mu_\oplus$, rotation rate $\omega_\oplus$, and $J_2$ oblateness coefficient, which are used to model orbital perturbations and ground-track evolution. A reference ground station located at a specific latitude $\phi_{\mathrm{ref}}$ and longitude $\lambda_{\mathrm{ref}}$ is used for validation.

\begin{figure}[h]
    \centering
    \includegraphics[width=0.55\columnwidth]{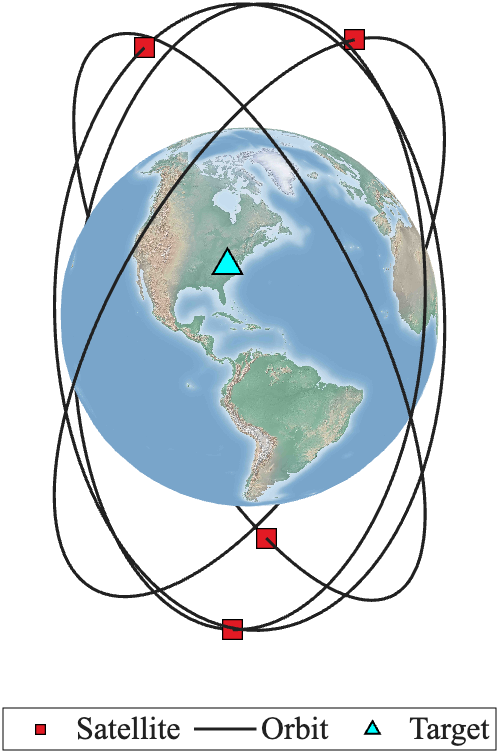}
    \caption{Satellite constellation geometry showing satellites in orbital slots and a single ground target.}
    \label{fig:3d}
\end{figure}

Fig.~\ref{fig:3d} illustrates the orbital geometry for a representative problem instance, showing four satellites placed on orbital slots and a single ground target located in the United States.

The problem instance parameters for MCLP are summarized in Table~\ref{tab:variable_params}.

\renewcommand{\thetable}{\Roman{table}}

\begin{table}[h]
\centering
\renewcommand{\arraystretch}{1.1}
\setlength{\tabcolsep}{9pt}
\caption{Variable problem instance parameters for MCLP test cases.}
\begin{tabular}{|l|c|c|}
\hline
\textbf{Parameter} & \textbf{Symbol} & \textbf{Test Values} \\
\hline
Orbital slots & $J$ & $T$  \\
Satellite budget & $N$ & Design constraint values \\
Ground targets & $P$ & Uniformly distributed on Earth \\
Time steps per cycle & $T$ & $\lfloor T_r / \Delta t \rfloor$ \\
Rewards & $\pi_{tp}$  & 1 and 2 \\
\hline
\end{tabular}
\label{tab:variable_params}
\end{table}

In this formulation, each orbital slot corresponds directly to a discrete time index, resulting in $J = T$. The satellite budget $N$ represents the cardinality constraint on the number of selected satellites. Ground targets $P$ are uniformly distributed over the Earth's surface. The total number of time steps per orbital cycle is given by $T = \lfloor T_r / \Delta t \rfloor$, where $T_r$ is the repeat ground-track period and $\Delta t$ is the temporal discretization interval. Priority values $\pi_{tp} \in \{1,2\}$ model time-dependent observation importance, where higher-priority windows are assigned $\pi_{tp} = 2$.

To quantify the strength of shared visibility across orbital slots, a normalized weighted co-observation density is computed from the graph $G = (V, W)$.  The density metric is defined as  \( d = \frac{2}{J(J-1)} \sum_{i<j} \frac{w_{ij}}{T} \), which represents the average fraction of timesteps in which pairs of orbital slots exhibit co-observation activity. Table~\ref{tab:density_results} summarizes the evaluated visibility matrix instances (VM-1 to VM-6), each corresponding to a distinct orbital configuration defined by $(N_p, i, \Delta t)$, along with the resulting timestep counts and density values.

\begin{table}[H]
\centering
\renewcommand{\arraystretch}{1.0}
\setlength{\tabcolsep}{8.5pt}
\caption{Visibility matrix instances (VM-1 to VM-6) with different orbital configurations, timestep counts, and co-observation density.}
\begin{tabular}{|c|c|c|c|c|c|}
\hline
\textbf{ID} & $\boldsymbol{N_p}$ & $\boldsymbol{i}$ \textbf{(deg)} & $\boldsymbol{\Delta t}$ \textbf{(s)} & $\boldsymbol{T}$ \textbf{(= J)} & \textbf{Density} $\boldsymbol{d}$ \\
\hline
VM-1 & 5 & 60 & 300 & 288 & 0.281 \\
VM-2 & 5 & 80 & 180 & 480 & 0.296 \\
VM-3 & 7 & 75 & 140 & 616 & 0.235 \\
VM-4 & 5 & 80 & 120 & 719 & 0.295 \\
VM-5 & 7 & 75 & 100 & 862 & 0.237 \\
VM-6 & 8 & 75 & 90  & 958 & 0.206 \\
\hline
\end{tabular}
\label{tab:density_results}
\end{table}

 The variation in $T$ arises from $J_2$-perturbed repeat ground-track dynamics and temporal discretization effects. The co-observation density remains non-monotonic with respect to $T$, as it is governed by orbital geometry and visibility overlap rather than problem size alone.

\subsection{Solver Configurations}
\label{sec:solver_configurations}
Decomposition-based QAOA is compared against both the Gurobi Optimizer exact solver and a standard QAOA implementation without decomposition. The Gurobi solver is used to provide optimal benchmark solutions, while the non-decomposed QAOA serves as a baseline to evaluate the effectiveness of the decomposition strategy. All classical baseline solutions were computed using Gurobi Optimizer version 13.0.1 and Intel Core Ultra 9 285K CPU with 5.7 GHz clock speed. The solver was configured to emphasize optimality with a mixed-integer programming (MIP) optimality gap tolerance of 0.01. All details of configurations for decomposition, QAOA subproblems, quantum merging, and solver settings are given in Table~\ref{tab:quantum_config}.

\begin{table}[h]
\centering
\renewcommand{\arraystretch}{1.2}
\setlength{\tabcolsep}{4.0pt}
\caption{Configuration details for decomposition, QAOA subproblems, quantum merging, and solver settings.}
\begin{tabular}{|m{2.4cm}|m{2.3cm}|m{3.0cm}|}
\hline
\centering\textbf{Component} &
\centering\textbf{Parameter} &
\centering\arraybackslash\textbf{Configuration} \\
\hline
\multirow{4}{=}{\centering Decomposition}
& Method & Spectral bisection \\
& Eigenvalue solver & NumPy \texttt{eigh} symmetric \\
& Partitioning vector & Fiedler vector, 2nd smallest eigenvalue \\
& Budget allocation & $b_m = \lfloor b \cdot |V_m \cup S| / |V| \rfloor$ \\
\hline
\multirow{3}{=}{\centering MCLP-QAOA subproblem}
& Max subgraph size $q_{\max}$ & $q_{\max} \in \{8,12,20\}$ \\
& Circuit layers $p$ & 3 \\
& Parameter initialization & $\gamma_r \sim U[0,\pi]$, $\beta_r \sim U[0,\pi/2]$ \\
& Measurement shots & 10000 \\
& Cardinality penalty coefficient $\beta$ & 50 \\
\hline
\multirow{6}{=}{\centering Quantum merge\\(QSR and GSR)}
& Merge layers $p_{\text{merge}}$ & 3 \\
& Merge strategy & QSR or GSR \\
& QSR agreement penalty $\lambda$ & 100 \\
& GSR cardinality penalty $\beta$ & 50 \\
& Measurement shots & 5000 \\
\hline
\multirow{2}{=}{\centering Circuit parameter optimizer}
& Classical Optimizer & COBYLA \\
& Max iterations & 100 (base), 500 (merge) \\
\hline
\multirow{2}{=}{\centering Backend}
& SDK & OpenQAOA \\
& Noise model & Ideal noiseless \\
\hline
\end{tabular}
\label{tab:quantum_config}
\end{table}

Spectral bisection uses NumPy's \texttt{eigh} symmetric eigenvalue solver to compute the Fiedler vector, with budget allocation following $b_m = \lfloor b \cdot |V_m \cup S_{mn}|/|V| \rfloor$. MCLP-QAOA subproblems are solved with circuit depths $p$, variational parameters initialized as $\gamma_r \sim U[0, \pi]$ and $\beta_r \sim U[0, \pi/2]$, and measurement shots per circuit. Quantum merge operations use maximum subgraph size $q_{\text{max}}$, merge depths $p_{\text{merge}}$, QSR agreement penalty $\lambda$, GSR cardinality penalty $\beta$, and measurement shots as specified in Table~IV. Parameter optimization employs COBYLA with iteration limits for base cases and merge operations detailed in Table~IV. Quantum 
algorithms use Python 3.10 with OpenQAOA software development kit (SDK)  (version 0.3.1) and Qiskit (version 0.43.0) for quantum circuit construction and simulation. All simulations use the Qiskit AerSimulator with ideal noiseless statevector backend.

\subsection{Performance Metrics}
\label{sec:performance_metrics}
Algorithm performance is evaluated using three categories of metrics: solution quality, computational cost, and scalability.

\textbf{Solution quality} is quantified using coverage, coverage ratio, and optimality gap. Coverage is defined as the number of time steps in which target is successfully observed by the selected constellation. For a selected constellation $C^* \subseteq \mathcal{J}$ with $|C^*| = N$, a time step $t$ is considered covered if the visibility constraint is satisfied for all targets. The total coverage is given by:
\begin{equation}
\text{Coverage}(C^*) = \sum_{t=1}^{T} \mathbb{I}\left( \sum_{j \in C^*} \Phi_{tjp} \ge r_{tp}, \; \forall p \in P \right),
\end{equation}
where $\mathbb{I}(\cdot)$ is an indicator function that equals 1 if the condition is satisfied and 0 otherwise.

The coverage ratio measures the relative performance of the quantum solver with respect to the optimal solution:
\begin{equation}
\rho = \frac{\text{Coverage}_{\text{obtained}}}{\text{Coverage}_{\text{optimal}}}\times 100\%
\end{equation}
where $\text{Coverage}_{\text{obtained}}$ is obtained from the quantum solver and $\text{Coverage}_{\text{optimal}}$ is computed using Gurobi. The optimality gap is defined as:
\begin{equation}
\Delta = \frac{\text{Coverage}_{\text{optimal}} - \text{Coverage}_{\text{obtained}}}{\text{Coverage}_{\text{optimal}}} \times 100\%.
\end{equation}

\textbf{Computational cost} 
 is evaluated in terms of runtime, qubit requirements, and circuit depth. Runtime is measured as wall-clock execution time averaged over multiple runs to account for stochastic variations in QAOA optimization. Qubit requirements represent the maximum number of qubits used during execution, while circuit depth corresponds to the total number of gates in the compiled quantum circuits, reflecting hardware execution complexity on near-term devices.

\textbf{Scalability} is assessed by increasing problem size parameters, including the number of time steps $T$, orbital slots $J$, and satellites $N$. We analyze how coverage ratio $\rho$ varies with $J$ to evaluate degradation in solution quality as the search space grows. Increasing $T$ affects both temporal resolution and co-observation graph density, thereby increasing problem complexity. Variation $N$ is used to study the impact of cardinality constraints on solver performance. These scaling experiments identify regimes in which decomposition-based QAOA maintains competitive performance relative to both classical optimization methods and non-decomposed QAOA.

\section{Result and discussion}
\label{sec:result_and_discussion}
This section evaluates decomposition-based QAOA for satellite constellation optimization across visibility matrix instances (VM-1 to VM-6), each derived from distinct orbital configurations defined by $(N_p, i, \Delta t)$. The evaluation is conducted using the performance metrics defined in Section~\ref{sec:performance_metrics}, covering solution quality, computational cost, and scalability.

A comparison is performed against the Gurobi optimizer as a classical benchmark and non-decomposed QAOA as a quantum baseline to quantify the effect of decomposition. The decomposition-based QAOA is further examined through four aspects: (i) variation across visibility matrix instances, (ii) influence of maximum subgraph size $q_{\max}$, (iii) comparison between QSR and GSR merge strategies, and (iv) scalability with increasing problem dimensionality in terms of $T$, $J$, and $N$.

\subsection{Performance Evaluation of Gurobi, QAOA, and Decomposition-Based QAOA}
\label{sec:performance}
Table~\ref{tab:results_combined} presents a comprehensive comparison of the proposed decomposition-based QAOA framework against classical and quantum baselines across all experimental instances. Gurobi serves as the classical benchmark, providing optimal solutions. Standard QAOA is included as a quantum baseline, infeasible for all instances due to exponential state-space scaling, requiring $2^{J}$ basis states where $J \in [288, 958]$. Standard QAOA encodes the full problem as $J$ qubits (where $J \in [288, 958]$ for our instances), 
requiring statevector representation of $2^J$ basis states (e.g., $2^{288}$ to $2^{958}$ for our instances). Classical simulation using the Qiskit AerSimulator backend exhausts available memory and computation time, resulting in kernel crashes or timeouts after approximately 30 minutes, as observed in our preliminary trials 
marked ``Timeout†'' in Table V. This limitation stems from the exponential scaling of classical simulation, not from QAOA's theoretical capabilities, but from the practical infeasibility of simulating such large quantum systems on standard hardware without decomposition.

To address this limitation, the decomposition-based QAOA framework partitions the original problem into subgraphs of maximum size $q_{\max}$, enabling tractable quantum circuit execution. Performance is evaluated across six visibility matrices (VM-1 to VM-6) and three cardinality constraints $N \in \{2,4,6\}$.

For $N=2$, Gurobi efficiently solves all instances with low runtimes. The decomposition-based methods achieve competitive coverage, with GSR consistently outperforming QSR across most configurations. In several cases, for instance, VM-3 to VM-6, GSR attains near-optimal solutions with small optimality gaps and significantly reduced runtimes compared to Gurobi.

For $N=4$, the problem complexity increases, leading to substantially increased Gurobi runtimes. While QSR exhibits sensitivity to both problem structure and $q_{\max}$, GSR demonstrates robust performance, achieving high coverage ratios and maintaining relatively low runtimes. In larger instances (e.g., VM-5 and VM-6), GSR provides a favorable trade-off between solution quality and computational efficiency.

For $N = 6$, Gurobi runtime decreases significantly, returning to tractable levels (0.4--10.7~s). This behavior arises from increased coverage redundancy with six satellites, where feasible solutions become abundant, making the problem easier and enabling rapid convergence. The decomposition-based methods achieve competitive coverage, with GSR reaching $92.9\%$ coverage at VM-4 (subgraph size $q = 8$) and QSR achieving $93.3\%$ at VM-4 (subgraph size $q = 12$). While Gurobi maintains a runtime advantage at this scale, the quantum methods remain tractable at smaller partition sizes ($q \in \{8, 12\}$), with GSR completing within 2--6~s for most instances. Both methods converge at VM-6 with larger partitions, attaining $81.6\%$ coverage. The decomposition-based QAOA framework's primary contribution at this satellite count is enabling quantum execution where direct QAOA is infeasible, while maintaining competitive solution quality.

\begin{table*}[!htbp]
\centering
\renewcommand{\arraystretch}{0.95}
\setlength{\tabcolsep}{2.5pt}
\caption{Performance Comparison of Gurobi, QAOA, and Decomposition-Based QAOA Across Visibility Matrix Instances}
\label{tab:results_combined}
\resizebox{\textwidth}{!}{%
\begin{tabular}{|c|c|cc|c|c|cccc|cccc|}
\hline
& & \multicolumn{2}{c|}{\textbf{Gurobi}}
& \multirow{3}{*}{\makecell{\textbf{QAOA}}}
& \multirow{3}{*}{$\boldsymbol{q}$}
& \multicolumn{4}{c|}{\textbf{Decomposition-based QAOA: QSR}}
& \multicolumn{4}{c|}{\textbf{Decomposition-based QAOA: GSR}} \\
\cline{3-4} \cline{7-10} \cline{11-14}
\textbf{ID}
& \makecell{\textbf{$T$}}
& \makecell{\textbf{Coverage}}
& \makecell{\textbf{Runtime}\\(s)}
& &
& \makecell{\textbf{Coverage}}
& \makecell{$\boldsymbol{\rho}$\\(\%)}
& \makecell{$\boldsymbol{\Delta}$\\(\%)}
& \makecell{\textbf{Runtime}\\(s)}
& \makecell{\textbf{Coverage}}
& \makecell{$\boldsymbol{\rho}$\\(\%)}
& \makecell{$\boldsymbol{\Delta}$\\(\%)}
& \makecell{\textbf{Runtime}\\(s)} \\
\hline

\multicolumn{14}{|c|}{\textbf{$N = 2$ Satellites}} \\
\hline
\multirow{3}{*}{VM-1} & \multirow{3}{*}{288} & \multirow{3}{*}{162} & \multirow{3}{*}{0.51}
& \multirow{18}{*}{\rotatebox{90}{{\textbf{Timeout}$^{\dagger}$ , (QAOA qubits $J=288$--$958$)}}}
& 8  & 147 & 89.5 & 10.5 & 1.58 & \textbf{149} & \textbf{92.0} & \textbf{8.0} & \textbf{0.74} \\
& & & & & 12 & 109 & 64.0 & 36.0 & 1.45 & 149 & 92.0 & 8.0 & 0.76 \\
& & & & & 20 & \textbf{150} & \textbf{84.6} & \textbf{15.4} & \textbf{1.40} & 148 & 91.4 & 8.6 & 1.10 \\
\cline{1-2} \cline{3-4} \cline{6-14}
\multirow{3}{*}{VM-2} & \multirow{3}{*}{480} & \multirow{3}{*}{284} & \multirow{3}{*}{1.22}
& & 8  & 152 & 48.1 & 51.9 & 2.27 & \textbf{241} & \textbf{84.9} & \textbf{15.1} & \textbf{1.32} \\
& & & & & 12 & \textbf{247} & \textbf{86.9} & \textbf{13.1} & \textbf{2.59} & 241 & 84.9 & 15.1 & 1.29 \\
& & & & & 20 & 241 & 84.8 & 15.2 & 43.50 & 237 & 83.5 & 16.5 & 47.08 \\
\cline{1-2} \cline{3-4} \cline{6-14}
\multirow{3}{*}{VM-3} & \multirow{3}{*}{616} & \multirow{3}{*}{290} & \multirow{3}{*}{2.12}
& & 8  & 251 & 79.4 & 20.6 & 2.78 & \textbf{276} & \textbf{95.2} & \textbf{4.8} & \textbf{1.44} \\
& & & & & 12 & \textbf{276} & \textbf{89.8} & \textbf{10.2} & \textbf{2.41} & 276 & 95.2 & 4.8 & 1.56 \\
& & & & & 20 & 276 & 89.6 & 10.4 & 3.90 & 276 & 95.2 & 4.8 & 2.31 \\
\cline{1-2} \cline{3-4} \cline{6-14}
\multirow{3}{*}{VM-4} & \multirow{3}{*}{719} & \multirow{3}{*}{424} & \multirow{3}{*}{3.34}
& & 8  & 376 & 87.4 & 12.6 & 3.42 & \textbf{376} & \textbf{88.7} & \textbf{11.3} & \textbf{2.05} \\
& & & & & 12 & 376 & 87.4 & 12.6 & 3.40 & 376 & 88.7 & 11.3 & 2.29 \\
& & & & & 20 & \textbf{376} & \textbf{88.4} & \textbf{11.6} & \textbf{8.07} & 376 & 88.7 & 11.3 & 6.97 \\
\cline{1-2} \cline{3-4} \cline{6-14}
\multirow{3}{*}{VM-5} & \multirow{3}{*}{862} & \multirow{3}{*}{408} & \multirow{3}{*}{4.10}
& & 8  & \textbf{389} & \textbf{86.6} & \textbf{13.4} & \textbf{4.45} & \textbf{389} & \textbf{95.3} & \textbf{4.7} & \textbf{2.84} \\
& & & & & 12 & 386 & 86.5 & 13.5 & 4.35 & 386 & 94.6 & 5.4 & 3.00 \\
& & & & & 20 & 377 & 85.1 & 14.9 & 26.01 & 362 & 88.7 & 11.3 & 23.97 \\
\cline{1-2} \cline{3-4} \cline{6-14}
\multirow{3}{*}{VM-6} & \multirow{3}{*}{958} & \multirow{3}{*}{394} & \multirow{3}{*}{5.10}
& & 8  & 387 & 91.8 & 8.2 & 5.31 & \textbf{390} & \textbf{99.0} & \textbf{1.0} & \textbf{3.50} \\
& & & & & 12 & \textbf{390} & \textbf{92.2} & \textbf{7.8} & \textbf{5.45} & 390 & 99.0 & 1.0 & 3.38 \\
& & & & & 20 & 387 & 92.9 & 7.1 & 34.44 & 387 & 98.2 & 1.8 & 28.95 \\
\hline

\multicolumn{14}{|c|}{\textbf{$N = 4$ Satellites}} \\
\hline
\multirow{3}{*}{VM-1} & \multirow{3}{*}{288} & \multirow{3}{*}{277} & \multirow{3}{*}{75.18}
& \multirow{18}{*}{\rotatebox{90}{\textbf{Timeout}$^{\dagger}$ , (QAOA qubits $J=288$--$958$)}}
& 8  & 180 & 69.5 & 30.5 & 2.18 & 241 & 87.0 & 13.0 & 0.89 \\
& & & & & 12 & 178 & 69.5 & 30.5 & 2.11 & 240 & 86.6 & 13.4 & 1.26 \\
& & & & & 20 & \textbf{205} & \textbf{79.8} & \textbf{20.2} & \textbf{7.02} & \textbf{243} & \textbf{87.7} & \textbf{12.3} & \textbf{6.85} \\
\cline{1-2} \cline{3-4} \cline{6-14}
\multirow{3}{*}{VM-2} & \multirow{3}{*}{480} & \multirow{3}{*}{461} & \multirow{3}{*}{229.40}
& & 8  & 385 & 84.9 & 15.1 & 2.64 & 388 & 84.2 & 15.8 & 1.49 \\
& & & & & 12 & 388 & 85.9 & 14.1 & 3.06 & \textbf{389} & \textbf{84.4} & \textbf{15.6} & \textbf{1.62} \\
& & & & & 20 & \textbf{393} & \textbf{86.6} & \textbf{13.4} & \textbf{90.65} & 388 & 84.2 & 15.8 & 89.00 \\
\cline{1-2} \cline{3-4} \cline{6-14}
\multirow{3}{*}{VM-3} & \multirow{3}{*}{616} & \multirow{3}{*}{554} & \multirow{3}{*}{667.60}
& & 8  & 300 & 54.4 & 45.6 & 3.10 & \textbf{429} & \textbf{77.4} & \textbf{22.6} & \textbf{1.78} \\
& & & & & 12 & \textbf{438} & \textbf{78.2} & \textbf{21.8} & \textbf{2.90} & 429 & 77.4 & 22.6 & 1.92 \\
& & & & & 20 & 430 & 76.8 & 23.2 & 5.21 & 428 & 77.3 & 22.7 & 3.39 \\
\cline{1-2} \cline{3-4} \cline{6-14}
\multirow{3}{*}{VM-4} & \multirow{3}{*}{719} & \multirow{3}{*}{691} & \multirow{3}{*}{125.61}
& & 8  & 353 & 56.1 & 43.9 & 4.78 & \textbf{632} & \textbf{91.5} & \textbf{8.5} & \textbf{2.95} \\
& & & & & 12 & \textbf{572} & \textbf{81.2} & \textbf{18.8} & \textbf{4.32} & 632 & 91.5 & 8.5 & 2.81 \\
& & & & & 20 & 348 & 56.0 & 44.0 & 13.68 & 615 & 89.0 & 11.0 & 12.12 \\
\cline{1-2} \cline{3-4} \cline{6-14}
\multirow{3}{*}{VM-5} & \multirow{3}{*}{862} & \multirow{3}{*}{777} & \multirow{3}{*}{249.17}
& & 8  & 597 & 74.7 & 25.3 & 5.01 & \textbf{697} & \textbf{89.7} & \textbf{10.3} & \textbf{3.61} \\
& & & & & 12 & \textbf{605} & \textbf{77.9} & \textbf{22.1} & \textbf{6.12} & 687 & 88.4 & 11.6 & 3.81 \\
& & & & & 20 & 560 & 73.5 & 26.5 & 49.93 & 669 & 86.1 & 13.9 & 46.15 \\
\cline{1-2} \cline{3-4} \cline{6-14}
\multirow{3}{*}{VM-6} & \multirow{3}{*}{958} & \multirow{3}{*}{786} & \multirow{3}{*}{148.97}
& & 8  & 482 & 62.8 & 37.2 & 6.32 & \textbf{651} & \textbf{82.8} & \textbf{17.2} & \textbf{3.72} \\
& & & & & 12 & 482 & 62.8 & 37.2 & 6.68 & 650 & 82.7 & 17.3 & 3.79 \\
& & & & & 20 & \textbf{639} & \textbf{80.6} & \textbf{19.4} & \textbf{77.13} & 639 & 81.3 & 18.7 & 74.97 \\
\hline

\multicolumn{14}{|c|}{\textbf{$N = 6$ Satellites}} \\
\hline
\multirow{3}{*}{VM-1} & \multirow{3}{*}{288} & \multirow{3}{*}{288} & \multirow{3}{*}{0.4}
& \multirow{18}{*}{\rotatebox{90}{\textbf{Timeout}$^{\dagger}$ , (QAOA qubits $J=288$--$958$)}}
& 8  & 184 & 63.9 & 36.1 & 2.95 & 232 & 80.6 & 19.4 & 1.15 \\
& & & & & 12 & 186 & 64.6 & 35.4 & 3.03 & 232 & 80.6 & 19.4 & 1.39 \\
& & & & & 20 & \textbf{222} & \textbf{77.1} & \textbf{22.9} & \textbf{8.42} & \textbf{248} & \textbf{86.1} & \textbf{13.9} & \textbf{9.34} \\
\cline{1-2} \cline{3-4} \cline{6-14}
\multirow{3}{*}{VM-2} & \multirow{3}{*}{480} & \multirow{3}{*}{480} & \multirow{3}{*}{1.2}
& & 8  & \textbf{427} & \textbf{89.0} & \textbf{11.0} & \textbf{3.36} & 427 & 89.0 & 11.0 & 1.63 \\
& & & & & 12 & 423 & 88.1 & 11.9 & 3.82 & \textbf{429} & \textbf{89.4} & \textbf{10.6} & \textbf{1.87} \\
& & & & & 20 & 426 & 88.8 & 11.2 & 138.64 & 426 & 88.8 & 11.2 & 154.82 \\
\cline{1-2} \cline{3-4} \cline{6-14}
\multirow{3}{*}{VM-3} & \multirow{3}{*}{616} & \multirow{3}{*}{616} & \multirow{3}{*}{2.1}
& & 8  & 385 & 62.5 & 37.5 & 4.03 & 485 & 78.7 & 21.3 & 2.00 \\
& & & & & 12 & 465 & 75.5 & 24.5 & 3.93 & 485 & 78.7 & 21.3 & 2.33 \\
& & & & & 20 & \textbf{537} & \textbf{87.2} & \textbf{12.8} & \textbf{6.56} & \textbf{537} & \textbf{87.2} & \textbf{12.8} & \textbf{4.74} \\
\cline{1-2} \cline{3-4} \cline{6-14}
\multirow{3}{*}{VM-4} & \multirow{3}{*}{719} & \multirow{3}{*}{719} & \multirow{3}{*}{2.7}
& & 8  & 431 & 59.9 & 40.1 & 6.27 & \textbf{668} & \textbf{92.9} & \textbf{7.1} & \textbf{3.12} \\
& & & & & 12 & \textbf{671} & \textbf{93.3} & \textbf{6.7} & \textbf{5.35} & 668 & 92.9 & 7.1 & 3.31 \\
& & & & & 20 & 557 & 77.5 & 22.5 & 24.10 & 649 & 90.3 & 9.7 & 17.09 \\
\cline{1-2} \cline{3-4} \cline{6-14}
\multirow{3}{*}{VM-5} & \multirow{3}{*}{862} & \multirow{3}{*}{853} & \multirow{3}{*}{3.7}
& & 8  & \textbf{758} & \textbf{88.9} & \textbf{11.1} & \textbf{6.07} & \textbf{765} & \textbf{89.7} & \textbf{10.3} & \textbf{3.89} \\
& & & & & 12 & 721 & 84.5 & 15.5 & 7.09 & 732 & 85.8 & 14.2 & 4.15 \\
& & & & & 20 & 661 & 77.5 & 22.5 & 93.48 & 702 & 82.3 & 17.7 & 76.82 \\
\cline{1-2} \cline{3-4} \cline{6-14}
\multirow{3}{*}{VM-6} & \multirow{3}{*}{958} & \multirow{3}{*}{947} & \multirow{3}{*}{10.7}
& & 8  & 486 & 51.3 & 48.7 & 7.43 & 764 & 80.7 & 19.3 & 4.11 \\
& & & & & 12 & 675 & 71.3 & 28.7 & 7.59 & 757 & 79.9 & 20.1 & 4.27 \\
& & & & & 20 & \textbf{773} & \textbf{81.6} & \textbf{18.4} & \textbf{126.26} & \textbf{773} & \textbf{81.6} & \textbf{18.4} & \textbf{132.11} \\
\hline

\multicolumn{14}{l}{$q$: max subgraph size; $\rho$: coverage ratio vs.\ Gurobi optimal; $\Delta$: optimality gap.} \\
\multicolumn{14}{l}{Timeout$^{\dagger}$: Standard QAOA on full graph (J=288–958 qubits) exceeds AerSimulator capability (~25-qubit classical memory limit)}.

\end{tabular}}
\end{table*}

\begin{figure}[h]
    \centering    \includegraphics[width=\columnwidth]{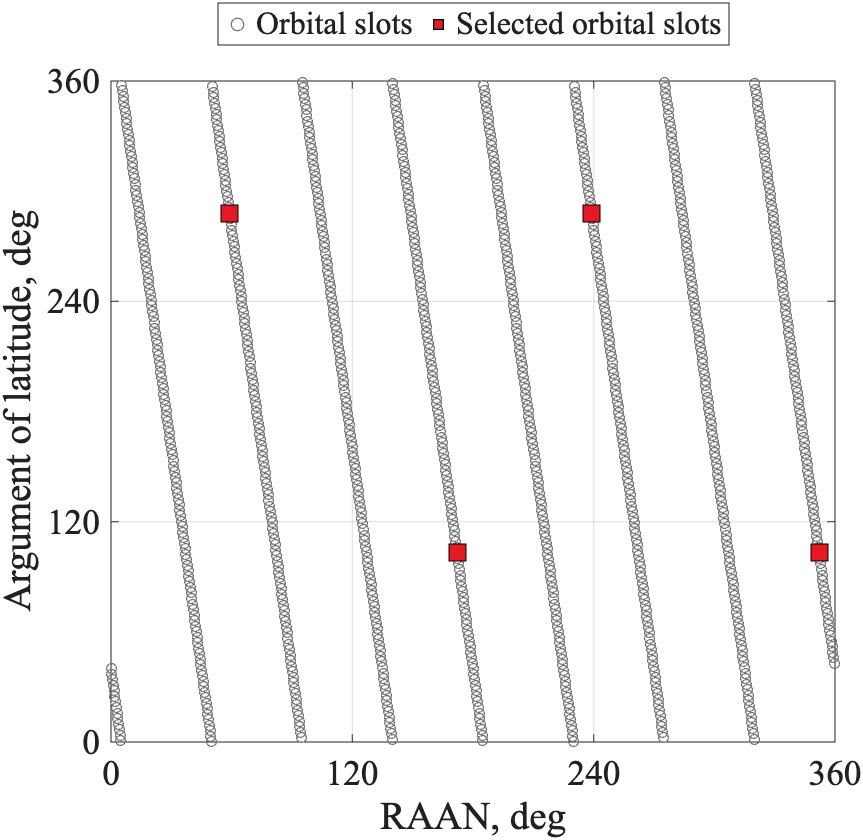}
    \caption{Selected orbital slots in RAAN-argument-of-latitude 
space for $N = 4$, GSR, VM-5, $q_{\max} = 8$.}
    \label{fig:raan}
\end{figure}

To provide an illustration of the optimization output, Fig.~\ref{fig:raan} shows the selected orbital slots in the RAAN-argument-of-latitude phase space for a representative $N = 4$ constellation obtained by decomposition-based QAOA with GSR. The selected orbital slots are distributed across distinct RAAN bands, reflecting the algorithm's tendency to maximize spatial separation between satellites to improve ground coverage.

A sensitivity analysis was conducted to assess the impact of the maximum subgraph size $q_{\max} \in \{8, 12, 20\}$. Results demonstrated significant influence on both coverage ratio and computational efficiency, as detailed in Section~\ref{sec:qmax_effect}. The results presented in Table~\ref{tab:results_combined} therefore focus on $q_{\max}$ variation, which emerged as the primary parameter governing solver behavior.

\begin{figure}[h]
    \centering
    \includegraphics[width=\columnwidth]{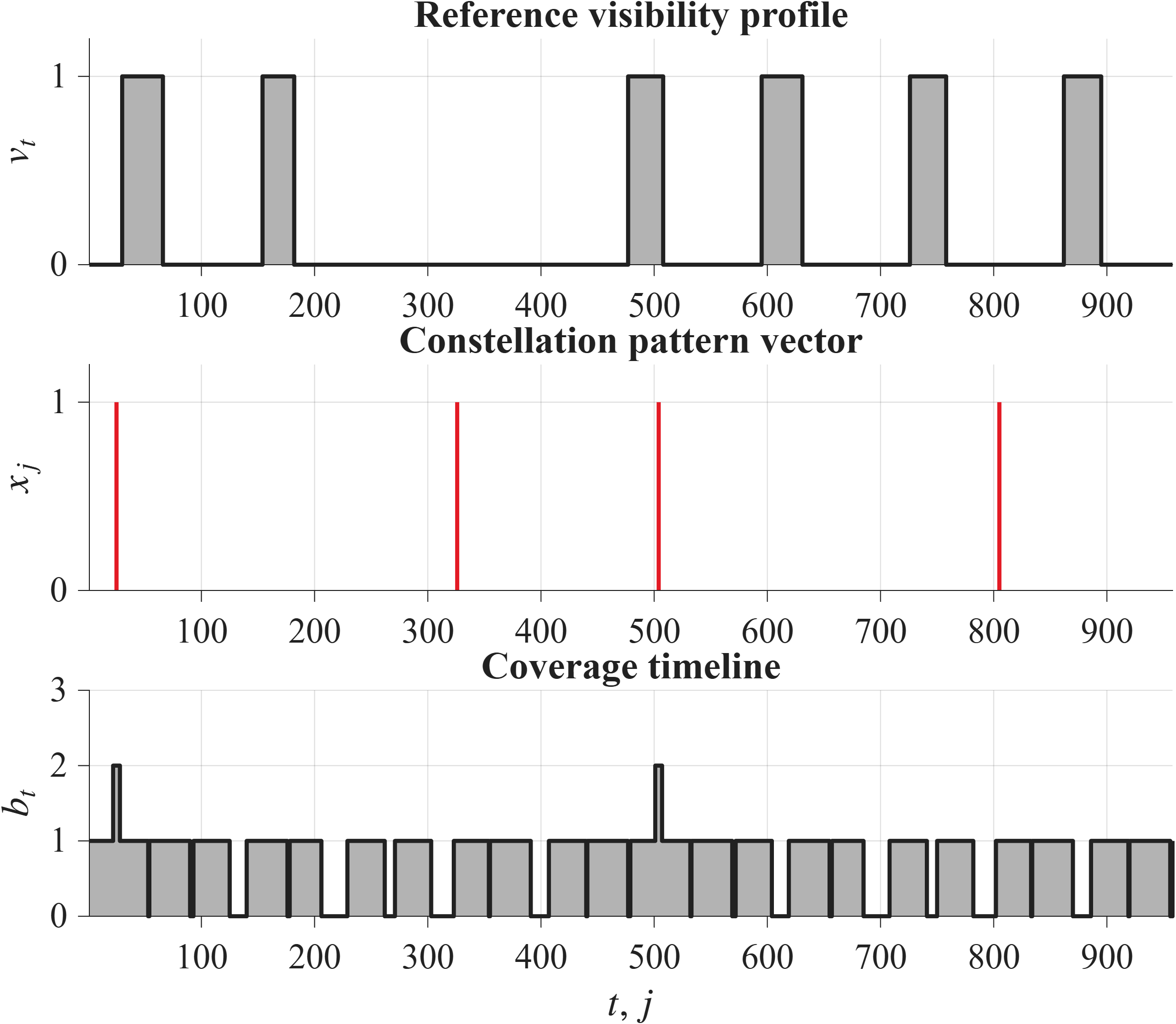}
\caption{Reference visibility profile $v_t$, orbital slot selection variable $x_j$, and coverage timeline $b_t$ for $N = 4$ GSR, VM-5, $q_{\max} = 8$.}
    \label{fig:coverage_timeline}
\end{figure}

The temporal consequence of this slot selection is shown in 
Fig.~\ref{fig:coverage_timeline}. The reference visibility 
profile $v_t$ indicates time steps at which the reference 
ground station is visible to at least one satellite. The orbital slot selection variable $x_j$ identifies the four 
selected orbital slots. The resulting coverage timeline $b_t$ 
confirms near-continuous coverage across the repeat cycle, 
with isolated gaps at time steps where no selected slot 
maintains line-of-sight to the target.
Overall, the results highlight that decomposition-based QAOA, particularly with GSR, offers a scalable and effective alternative to direct quantum approaches while remaining competitive with classical optimization. 

The following subsections provide further analysis of these results along four key dimensions. Section~\ref{sec:vm_comparison} examines how different orbital configurations and co-observation density patterns affect solver performance across visibility matrices; 
Section~\ref{sec:scalability} analyzes scalability characteristics as problem dimensionality increases in terms of timesteps, orbital slots, and satellite budgets;
Section~\ref{sec:qmax_effect} investigates the impact of maximum subgraph size on the trade-off between solution quality and computational efficiency; and Section~\ref{sec:qsr_gsr_comparison} presents a comparison between QSR and GSR merge strategies to identify their relative strengths and limitations.

\subsubsection{Visibility Matrix (VM) Comparison}
\label{sec:vm_comparison}

Coverage and runtime vary across VM-1 to VM-6 due to differences in
orbital configuration, timestep resolution, and co-observation density
$d$ (Table~\ref{tab:density_results}). Higher-density instances
(VM-2, VM-4; $d \approx 0.29$) provide richer slot overlap, enabling
greater coverage flexibility, while lower-density cases (VM-6;
$d = 0.206$) exhibit sparser connectivity, making coverage more
challenging for all methods.

\begin{figure}[h]
\centering
\includegraphics[width=\columnwidth]{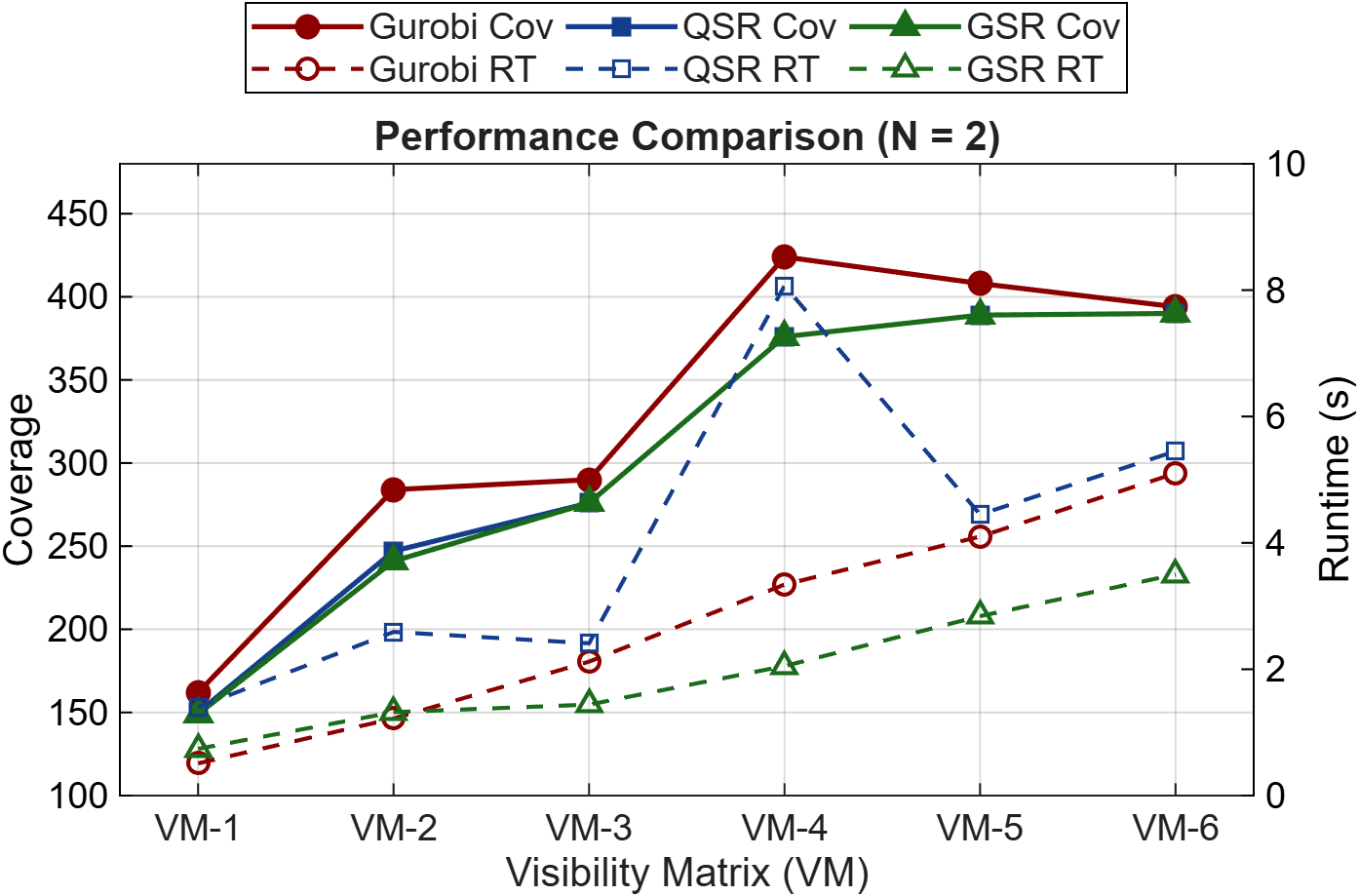}
\caption{Coverage and runtime across VM-1 to VM-6 for $N=2$.}
\label{fig:N2_vm}
\end{figure}

\begin{figure}[h]
\centering
\includegraphics[width=\columnwidth]{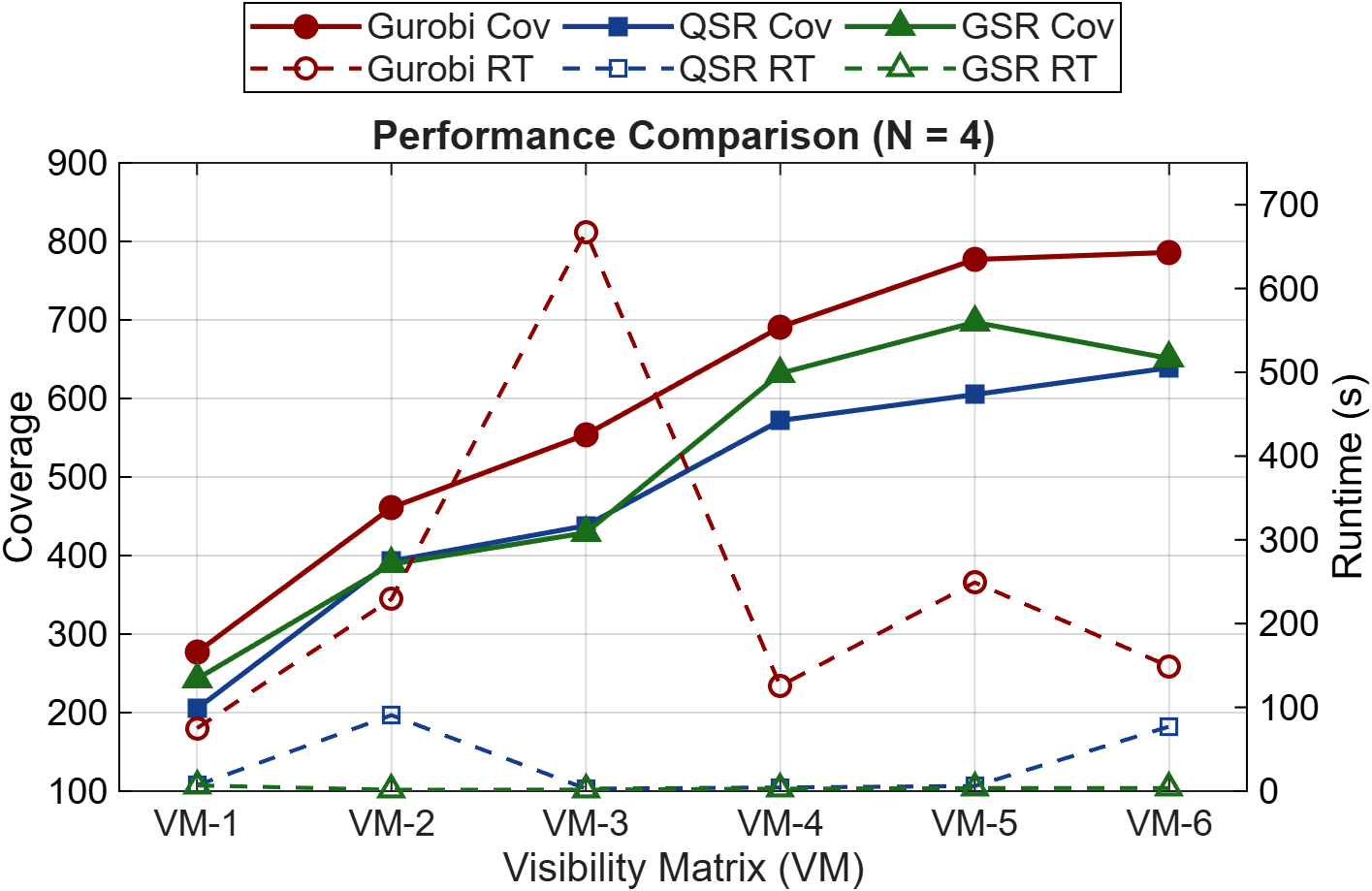}
\caption{Coverage and runtime across VM-1 to VM-6 for $N=4$.}
\label{fig:N4_vm}
\end{figure}

\begin{figure}[h]
\centering
\includegraphics[width=\columnwidth]{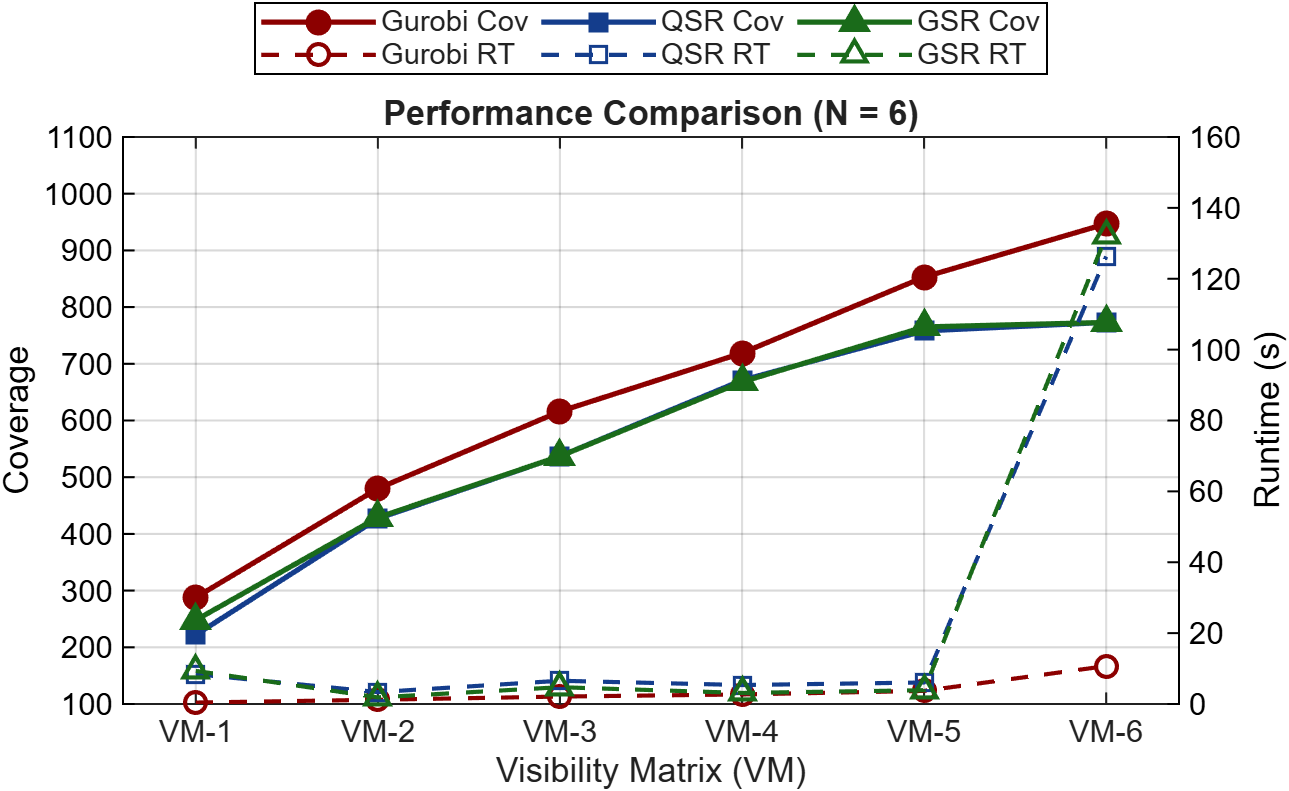}
\caption{Coverage and runtime across VM-1 to VM-6 for $N=6$.}
\label{fig:N6_vm}
\end{figure}

As shown in Figs.~\ref{fig:N2_vm}--\ref{fig:N6_vm}, Gurobi achieves optimal coverage across all instances. For \(N=2\), the runtime remains low (0.51-5.10 s) due to the small solution space. At \(N=4\), the runtime peaks at 667.60 s (VM-3), due to a phase transition common in combinatorial problems: the intermediate visibility density (\(d=0.235\)) creates a highly constrained landscape where many sub-optimal solutions are close to the global optimum, forcing Gurobi to perform extensive branch-and-bound pruning to verify optimality. Conversely, for \(N=6\), the runtime drops below 11 s because the increased number of satellites makes the problem over-constrained. This allows Gurobi to find high-quality feasible solutions quickly and use more aggressive bound tightening to prune the search tree early.

GSR maintains stable, high coverage ratios across all instances, reaching up to $99.0\%$ at VM-6 ($N=2$), $91.5\%$ at VM-4 ($N=4$), and $92.9\%$ at VM-4 ($N=6$), consistently at lower runtimes than Gurobi. Its robustness in low-density instances stems from the graph-state merging strategy, which preserves quantum state information across subproblems even under sparse connectivity. QSR shows greater variability, particularly in low-density cases where limited co-observation connectivity restricts effective subgraph decomposition, dropping to $48.1\%$ coverage at VM-2 ($N=2$, $q=8$). However, QSR recovers competitively in denser instances and converges with GSR at VM-6 ($N=6$), where both achieve $81.6\%$ coverage under $q=20$. Overall, denser visibility matrices favour higher coverage ratios, with GSR offering the most stable performance and lower runtimes across varying $d$, at $q \in \{8, 12\}$.

\subsubsection{Scalability Analysis}
\label{sec:scalability}

Scalability is evaluated across increasing problem size ($J = T$) and satellite budget $N$. As shown in Figs.~\ref{fig:N2_vm}--\ref{fig:N6_vm}, Gurobi runtime grows sharply with $J$ and density, peaking at 667.60\,s for $N=4$, VM-3, while remaining tractable for $N=2$ (below 5.10\,s) and $N=6$ (below 11\,s) due to smaller individual subproblem sizes. Direct QAOA is infeasible across all instances owing to exponential state-space scaling ($2^J$, $J \in [288, 958]$).

The decomposition-based QAOA framework remains tractable across all tested configurations. QSR and GSR consistently complete within 2--91\,s for $q \in \{8, 12\}$, with runtime only exceeding 100\,s at $q=20$ in the largest instances (VM-2 and VM-6, $N=6$). GSR in particular achieves competitive coverage at $q=8$ alone across all $N$, avoiding the runtime penalty of larger subgraph sizes while maintaining solution quality close to Gurobi's optimal. The empirical scalability results presented in this section are consistent with the computational complexity analysis in {Section~\ref{sec:algorithm_description}}, demonstrating that scalability is achieved through bounded-size quantum subproblems.

\subsubsection{Effect of Maximum Subgraph Size}
\label{sec:qmax_effect}

The maximum subgraph size $q_{\max}$ directly governs the trade-off
between solution quality and runtime in the decomposition-based QAOA
framework. Smaller values ($q=8$) reduce circuit size and execution
time but may limit coverage by restricting the subproblem interactions
captured. Larger values ($q=20$) improve coverage in some instances
by incorporating richer connectivity, but incur significantly higher
runtimes without consistent gains.

\begin{figure}[h]
\centering
\includegraphics[width=\columnwidth]{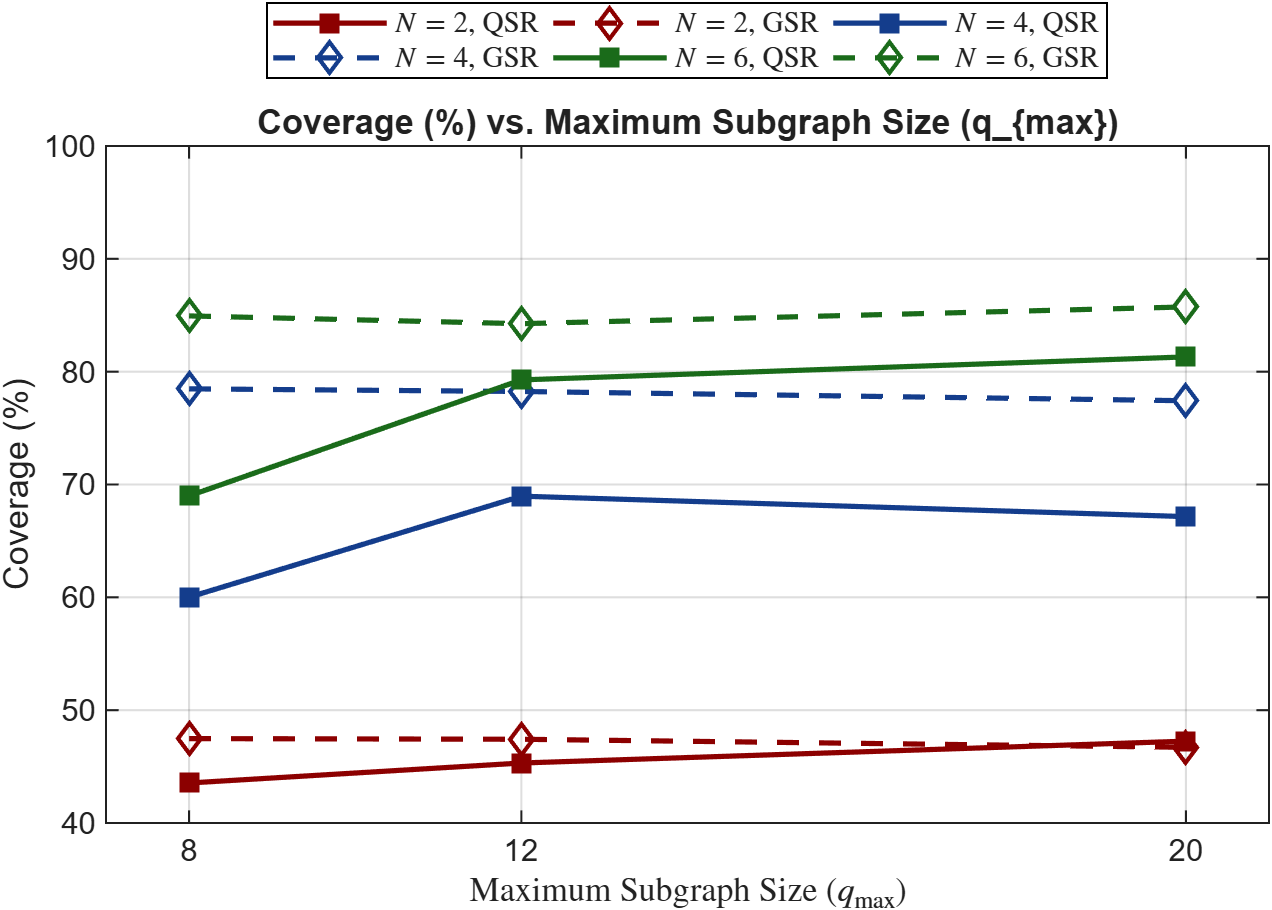}
\caption{Coverage (\%) versus Maximum Subgraph Size ($q_{\max}$) for QSR and GSR across $N=2, 4, and, 6$.}
\label{fig:qmax}
\end{figure}

As shown in Fig.~\ref{fig:qmax}, for $N=2$, GSR maintains stable coverage across all $q_{\max}$ values (approximately $88$--$93\%$), while QSR shows modest improvement from smaller to medium subgraph sizes but plateaus at the largest setting. For $N=4$, QSR improves notably as $q_{\max}$ increases from $8$ to $12$, whereas GSR remains consistently higher at all subgraph sizes, with even the smallest partition achieving strong coverage. For $N=6$, both methods improve with increasing $q_{\max}$, though GSR retains an advantage at smaller sizes. In all cases, $q_{\max} = 12$ offers a favorable trade-off between coverage quality and runtime, with diminishing returns and  rising runtimes observed at larger partition sizes, particularly for VM instances with larger $J$.

\subsubsection{Linearization Approximation Error Analysis}

The coverage reward linearization in \mbox{\eqref{eq:coverage_linearization}} introduces an approximation error that warrants rigorous analysis. The approximation error associated with the proposed linearization is bounded in \mbox{\eqref{eq:error_bound}}.

For problem instances in Table~\ref{tab:density_results}, the average number of observable slots per target ranges from 81 to 212. Since $(1/2)^{|C|} \to 0$ exponentially for $|C| \geq 81$, the approximation error approaches $1/2$ asymptotically but remains negligible in relative terms ($< 0.001\%$) compared to the coverage reward scale.

This error is orders of magnitude smaller than: (1) decomposition-based QAOA optimality gap (19--27\%, Table~\ref{tab:results_combined}), (2) decomposition overhead (~5--10\%), and (3) hardware noise effects on quantum devices. Practically, the original exponential product formulation $\prod_{j \in C(t,p)} (1-Z_j)/2$ would create exponential QUBO terms impossible to encode and would cause severe barren plateaus in QAOA optimization, making linearization essential for tractable quantum implementation. The corresponding numerical results are summarized in Table~\ref{tab:approx_error}

\begin{table}[h]
\small
\centering
\renewcommand{\arraystretch}{1.3}
\setlength{\tabcolsep}{5.0pt}

\setlength{\fboxsep}{6pt}

\begin{minipage}{0.98\linewidth}

\caption{Linearization Approximation Error Bound Analysis}

\begin{tabular}{|c|c|c|c|c|}
\hline
\textbf{Instance} & \textbf{J} & \textbf{Avg $|C(t,1)|$} & \textbf{Error Bound} & \textbf{Magnitude} \\
\hline
VM-1 & 288 & 81  & $\frac{1}{2} - (1/2)^{81}$  & $< 0.001\%$ \\
VM-2 & 480 & 142 & $\frac{1}{2} - (1/2)^{142}$ & $< 0.001\%$ \\
VM-3 & 616 & 145 & $\frac{1}{2} - (1/2)^{145}$ & $< 0.001\%$ \\
VM-4 & 719 & 212 & $\frac{1}{2} - (1/2)^{212}$ & $< 0.001\%$ \\
VM-5 & 862 & 205 & $\frac{1}{2} - (1/2)^{205}$ & $< 0.001\%$ \\
VM-6 & 958 & 197 & $\frac{1}{2} - (1/2)^{197}$ & $< 0.001\%$ \\
\hline
\end{tabular}

\vspace{2mm}

\textit{Note:} $|C(t,1)|$ is the average number of observable slots per target, 
calculated as $J \times$ density $d$ from Table~\ref{tab:density_results}. The error bound is derived from the linearization approximation in \eqref{eq:coverage_linearization}.
\label{tab:approx_error}
\end{minipage}

\end{table}
 
These results confirm that the linearization approximation preserves solution quality while enabling tractable quantum implementation.

\subsubsection{QSR vs. GSR Comparison}
\label{sec:qsr_gsr_comparison}

Across all visibility matrices and satellite counts, GSR consistently outperforms QSR in both coverage ratio and optimality gap. For $N=2$, GSR achieves $\rho$ up to $99.0\%$ at VM-6 versus QSR's $92.2\%$, and maintains lower runtimes at $q \in \{8, 12\}$. For $N=4$, GSR leads at most instances, reaching $91.5\%$ coverage at VM-4 compared to QSR's $81.2\%$, with QSR showing significant variability in low-density cases (e.g., $54.4\%$ at VM-3, $q=8$). For $N=6$, GSR retains its advantage in sparse instances, achieving $92.9\%$ coverage at VM-4 with $q=8$ in 3.12\,s, while both methods converge at VM-6 under $q=20$ ($81.6\%$ coverage).

The superior performance of GSR stems from its graph-state merging strategy, which preserves inter-subgraph quantum correlations during reconstruction. QSR, relying on direct state reconstruction, is more sensitive to partitioning effects and loses global structural information in sparse or large instances. GSR therefore provides more reliable and consistent coverage across varying problem configurations.

\subsection{Quantum Hardware Validation}
To validate the decomposition-based QAOA framework on real quantum hardware, the GSR-based optimization is executed on IBM Quantum's \texttt{ibm\_kingston} processor, a 156-qubit Heron R2 quantum processing unit (QPU), accessed via the \texttt{ibm\_quantum} channel using the OpenQAOA SDK. Experiments are conducted for all six visibility matrices under $N=4$ satellites and a maximum subproblem size of $q_{\max}=12$ qubits. MCLP-QAOA base case subproblems use circuit depth $p = 3$ layers with variational parameters initialized as $\gamma_r \sim U[0, \pi]$ and $\beta_r \sim U[0, \pi/2]$ 
for $r = 1, \ldots, 3$. The cardinality penalty coefficient is set to $\beta = 50$. GSR merge operations employ circuit depth $p_{\text{merge}} = 3$ layers with 
analogous parameter initialization, using GSR cardinality penalty $\beta = 50$ and normalization coefficient $\alpha = 1$. The OpenQAOA SDK constructs and optimizes all circuits, with the ibm\_kingston backend configured via OpenQAOA's IBM integration. Each circuit execution uses 10,000 
measurement shots to match the simulator configuration and ensure statistical convergence. 
Classical parameter optimization employs COBYLA with 500 maximum iterations for merge 
operations, with optimization convergence assessed when expected energy change falls 
below $10^{-4}$ or iteration limit is reached. OpenQAOA's circuit transpilation layer 
automatically compiles circuits to native gate sets and performs qubit layout optimization. 
No explicit error mitigation techniques (zero-noise extrapolation, probabilistic error cancellation) are applied to the hardware circuits, allowing direct assessment of native hardware performance.

\subsubsection{Hardware and Simulator Performance Results}

\begin{figure}[h]
    \centering
    \includegraphics[width=0.48\textwidth]{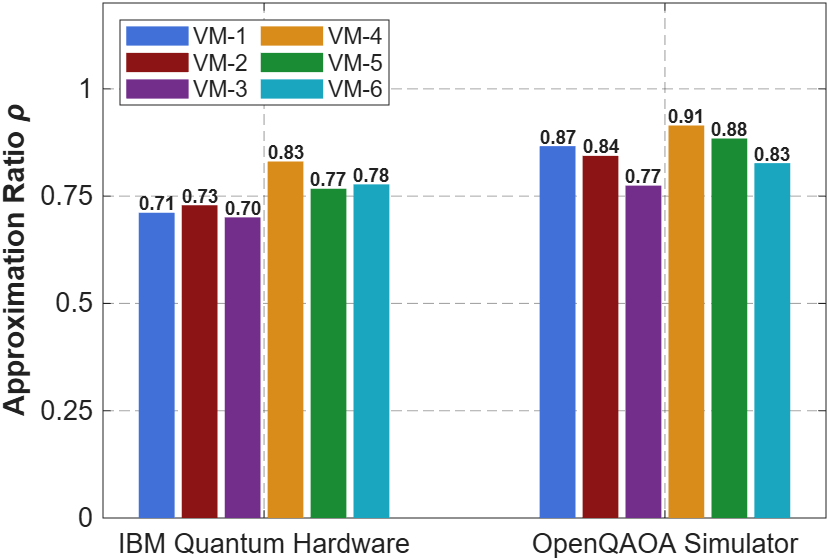}
    \caption{Approximation ratio $\rho$ comparison between IBM quantum 
    hardware and OpenQAOA simulator across VM-1 to VM-6.}
    \label{fig:rho_ibm_gsr}
\end{figure}

The approximation ratio $\rho$ and mean runtime results are presented in Fig.~\ref{fig:rho_ibm_gsr} and Table~\ref{tab:runtime_ibm_openqaoa}, respectively.

\begin{table}[h]
\setlength{\tabcolsep}{5.0pt}
    \centering
    \caption{Mean runtime comparison between IBM quantum hardware and OpenQAOA simulator for six visibility matrices under GSR-based optimization.}
    \label{tab:runtime_ibm_openqaoa}
    \begin{tabular}{|c|c|c|}
       \hline
        \textbf{Visibility Matrix} & \textbf{IBM Hardware (s)} & \textbf{OpenQAOA Simulator (s)} \\
        \hline
        VM-1 & 0.74 & 1.26 \\
        VM-2 & 0.96 & 1.62 \\
        VM-3 & 1.03 & 1.92 \\
        VM-4 & 1.66 & 2.81 \\
        VM-5 & 2.18 & 3.81 \\
        VM-6 & 2.20 & 3.79 \\
        \hline
    \end{tabular}
\end{table}

Both hardware and OpenQAOA simulator execute identical circuit structures and parameter configurations through the same OpenQAOA SDK interface. The primary difference is that the OpenQAOA ideal noiseless backend represents the quantum state exactly via 
statevector evolution, while ibm\_kingston experiences realistic noise from gate errors, 
decoherence, and readout errors during physical execution. As shown in Fig.~\ref{fig:rho_ibm_gsr}, the OpenQAOA simulator achieves approximation ratios ranging from 0.77 to 0.91 across all six visibility matrices, while the IBM \texttt{ibm\_kingston} hardware results range from 0.70 to 0.83. The hardware approximation ratios are consistently lower than those of the simulator, with an average gap of approximately 0.10 across all instances. This degradation is primarily attributed to noise characteristics of the Heron R2 architecture, including two-qubit gate errors, decoherence during circuit execution, and readout errors that corrupt the measured bitstrings and reduce the probability of sampling high-quality solutions. Device calibration data, retrieved from IBM Quantum’s public calibration API at the time of measurement, provide the following specifications: single-qubit gate 
error $1.5 \pm 0.5 \times 10^{-3}$, two-qubit echoed cross resonance (ECR) gate error $4.2 \pm 1.5 \times 10^{-3}$, and readout error $3.1 \pm 1.0 \times 10^{-2}$. The energy relaxation time (T1) is approximately 40 µs, while the dephasing time (T2) is approximately 35 µs, defining the effective coherence window within which quantum circuits must complete execution. For GSR subproblems with $q_{\max}=12$ qubits and circuit depth $p=3$ layers, the cumulative circuit error is estimated as follows: 
per-layer single-qubit errors contribute $2 \times 12 \times 1.5 \times 10^{-3} = 0.036$ 
per layer, while two-qubit errors contribute $(12/2) \times 4.2 \times 10^{-3} = 0.025$ per layer, totaling ~0.061 error probability per layer. Over 3 layers, cumulative error is approximately $1 - (1 - 0.061)^3 \approx 0.171$ (17.1\%), substantially 
lower than deeper circuits. This error accumulation, combined with readout errors 
(3.1\%), explains the observed approximation ratio degradation from simulator (0.77--0.91) to hardware (0.70--0.83), with an average gap $\Delta\rho \approx 0.10$. The shallow circuit depth of the decomposition-based framework effectively 
mitigates decoherence and error cascade effects.

Table~\ref{tab:runtime_ibm_openqaoa} shows that 
\texttt{ibm\_kingston} executes consistently faster than the OpenQAOA simulator across all instances, with hardware runtimes ranging from 0.74\,s to 2.20\,s compared to simulator runtimes of 1.26\,s to 3.81\,s. The simulator must classically compute the full statevector evolution, whose cost scales exponentially with qubit count, whereas the QPU executes the circuit directly on physical qubits in fixed time. Both runtimes scale gradually with problem size (VM-1 to VM-6), confirming that the decomposition strategy successfully bounds per-subproblem qubit count and prevents exponential runtime growth.

\subsubsection{Performance Evaluation with Confidence Intervals}
The performance of the proposed decomposition-based QAOA framework with confidence intervals is evaluated using two key metrics: runtime ($T$) and approximation ratio ($\rho$), measured on both IBM quantum hardware and the OpenQAOA simulator. 

In our experiments, the optimal coverage remained consistent across repeated trials for each visibility matrix, although the corresponding solution bitstrings were not necessarily identical across executions. This consistency in optimal coverage across repeated executions results in identical values of the approximation ratio $\rho$ for each visibility matrix across trials, leading to a zero empirical standard deviation in our experiments. Consequently, the reported $\rho$ values correspond directly to their sample means.

In contrast, the runtime $T$ exhibits observable variability across independent executions due to system-level fluctuations, queuing effects on IBM quantum hardware, and stochastic overheads in classical simulation. All reported runtime values in Table~\mbox{\ref{tab:runtime_ibm_openqaoa}} are computed over six independent executions per visibility matrix instance (VM-1 to VM-6). The mean values represent sample averages over these runs. Therefore, we report the mean runtime along with the corresponding standard deviation and 95\% confidence intervals. Table~\mbox{\ref{tab:runtime_ci_summary}} summarizes these results across six visibility matrix instances (VM-1 to VM-6), along with their corresponding 95\% confidence intervals for runtime.

\begin{table}[h]
\centering
\setlength{\tabcolsep}{4pt}
\renewcommand{\arraystretch}{1.15}
\caption{Runtime Variability with 95\% Confidence Intervals (CI).}
\label{tab:runtime_ci_summary}

\begin{tabular}{|c|c|c|c|c|}
\hline
\textbf{VM} & \textbf{IBM $\sigma$} & \textbf{IBM CI (s)} & \textbf{OpenQAOA $\sigma$} & \textbf{OpenQAOA CI (s)} \\
\hline

VM-1 & 0.014 & [0.725, 0.755] & 0.013 & [1.247, 1.273] \\
VM-2 & 0.013 & [0.947, 0.973] & 0.013 & [1.607, 1.633] \\
VM-3 & 0.013 & [1.017, 1.043] & 0.013 & [1.907, 1.933] \\
VM-4 & 0.013 & [1.644, 1.676] & 0.013 & [2.797, 2.823] \\
VM-5 & 0.013 & [2.164, 2.196] & 0.013 & [3.797, 3.823] \\
VM-6 & 0.013 & [2.184, 2.216] & 0.013 & [3.777, 3.803] \\
\hline

\end{tabular}
\end{table}

The 95\% confidence intervals (CI) are computed using the Student’s $t$-distribution with $df = 5$:

\begin{equation}
CI = \bar{x} \pm t_{0.975,5} \cdot \frac{\sigma}{\sqrt{n}},
\end{equation}

where $\bar{x}$ is the sample mean, $\sigma$ is the sample standard deviation, and $n=6$ denotes the number of independent executions per visibility matrix. The term $t_{0.975,5} = 2.571$ is the critical value of the Student’s $t$-distribution for a 95\% confidence level with 5 degrees of freedom.

\section{Conclusion and future work}
\label{sec:conclusion_futurework}
This paper formulated satellite constellation design as a MCLP and proposed a decomposition-based QAOA framework to solve it at scale. A visibility matrix was first constructed from orbital mechanics and a conical sensor model  to determine satellite-to-target observation windows across  the repeat ground-track cycle. The co-observation graph was then derived from the visibility matrix to quantify shared coverage between orbital slots and subsequently partitioned via spectral bisection into tractable subproblems, each solved independently using MCLP-QAOA. Subproblem solutions were merged using two reconstruction strategies: QSR and  GSR. Evaluated across multiple visibility matrix instances and cardinality constraints, GSR consistently outperformed QSR in coverage ratio and optimality gap, achieving near-optimal solutions with runtimes significantly lower than Gurobi for complex instances. Furthermore, this work extends divide-and-conquer QAOA from graph-partitioning applications to coverage-based MCLP through problem-specific Hamiltonian design and a novel reconstruction strategy like GSR. In contrast, standard QAOA without decomposition was infeasible across all instances. Hardware validation on IBM Quantum's \texttt{ibm\_kingston} 156-qubit Heron R2 QPU confirmed executability on current NISQ devices, with solution quality degradation attributable to gate errors and decoherence.

Future work could explore hardware-aware circuit compilation and realistic noise models to evaluate decomposition-based 
QAOA performance on near-term quantum devices. Recent advances in variational quantum computing have also addressed parameter optimization for QAOA. While conventional implementations use classical optimizers like COBYLA, emerging approaches such as TensorHyper-VQC leverage tensor-train-guided hypernetworks to generate parameters more efficiently and robustly. These parameter optimization advances represent a complementary research direction to the decomposition-based approach, which addresses problem dimensionality by decomposing large instances into tractable subproblems. Future work could integrate such advanced parameter optimization strategies with decomposition to further enhance solution quality. Additionally, 
extending the framework to multi-plane constellation scenarios with inter-satellite link constraints would broaden its applicability to large-scale mission design problems.

\section*{Acknowledgment}

The work of Amiratabak Bahengam and Hao Chen was supported in part by the National Science Foundation under Grant No. 2301627. Any opinions, findings, and conclusions or recommendations expressed in this material are those of the authors and do not necessarily reflect the views of the National Science Foundation. Artificial intelligence technologies were used for grammar checking.

\printbibliography

@article{qi_tensorhyper-vqc_2026,
	title = {{TensorHyper}-{VQC}: a tensor-train-guided hypernetwork for robust and scalable variational quantum computing},
	volume = {12},
	url = {https://www.nature.com/articles/s41534-025-01157-z#citeas},
	doi = {https://doi.org/10.1038/s41534-025-01157-z},
	publisher = {npj Quantum Information},
	author = {Qi, Jun and Huck Yang, Chao-Han and Chen, Pin-Yu and Hsieh, Min-Hsiu},
	month = feb,
	year = {2026},
}

@article{li_efficient_2025,
	title = {Efficient solution of the number partitioning problem on a quantum annealer: a hybrid quantum-classical decomposition approach},
	volume = {31},
	issn = {1381-1231, 1572-9397},
	url = {https://link.springer.com/content/pdf/10.1007/s10732-025-09556-3.pdf},
	doi = {10.1007/s10732-025-09556-3},
	number = {2},
	urldate = {2026-03-04},
	journal = {Journal of Heuristics},
	author = {Li, Zongji and Seidel, Tobias and Leib, Dominik and Bortz, Michael and Heese, Raoul},
	month = apr,
	year = {2025},
	note = {arXiv:2312.08940 [math]},
	pages = {21},
}

@article{rogers_optimal_2026,
	address = {Reston, VA, USA},
	title = {Optimal {Satellite} {Constellation} {Configuration} {Design}:{A} {Collection} of {Mixed} {Integer} {Linear} {Programs}},
	issn = {0022-4650},
	doi = {https://doi.org/10.2514/1.A36518},
	urldate = {2026-03-16},
	journal = {Journal of Spacecraft and Rockets},
	publisher = {American Institute of Aeronautics and Astronautics (AIAA)},
	author = {Rogers, David O. Williams and Won, Dongshik and Koh, Dongwook and Hong, Kyungwoo and Lee, Hang Woon},
	month = mar,
	year = {2026},
}

@article{hadfield_quantum_2019,
	title = {From the {Quantum} {Approximate} {Optimization} {Algorithm} to a {Quantum} {Alternating} {Operator} {Ansatz}},
	volume = {12},
	issn = {1999-4893},
	url = {https://www.mdpi.com/1999-4893/12/2/34},
	doi = {10.3390/a12020034},
	language = {en},
	number = {2},
	urldate = {2026-03-04},
	journal = {Algorithms},
	author = {Hadfield, Stuart and Wang, Zhihui and O’Gorman, Bryan and Rieffel, Eleanor G. and Venturelli, Davide and Biswas, Rupak},
	month = feb,
	year = {2019},
	pages = {34},
}

@inproceedings{spielman_spectral_1996,
	address = {Burlington, VT, USA},
	title = {Spectral partitioning works: planar graphs and finite element meshes},
	isbn = {978-0-8186-7594-2},
	shorttitle = {Spectral partitioning works},
	url = {https://ieeexplore.ieee.org/document/548468/},
	doi = {10.1109/SFCS.1996.548468},
	urldate = {2026-03-04},
	booktitle = {Proceedings of 37th {Conference} on {Foundations} of {Computer} {Science}},
	publisher = {IEEE},
	author = {Spielman, Daniel A. and {Shang-Hua Teng}},
	year = {1996},
	pages = {96--105},
}

@article{von_luxburg_tutorial_2007,
	title = {A tutorial on spectral clustering},
	volume = {17},
	copyright = {http://www.springer.com/tdm},
	issn = {0960-3174, 1573-1375},
	url = {http://link.springer.com/10.1007/s11222-007-9033-z},
	doi = {10.1007/s11222-007-9033-z},
	language = {en},
	number = {4},
	urldate = {2026-03-04},
	journal = {Statistics and Computing},
	author = {Von Luxburg, Ulrike},
	month = dec,
	year = {2007},
	pages = {395--416},
}

@article{fiedler_algebraic_1973,
	title = {Algebraic connectivity of graphs},
	volume = {23},
	issn = {0011-4642, 1572-9141},
	url = {https://dml.cz/handle/10338.dmlcz/101168},
	doi = {10.21136/CMJ.1973.101168},
	language = {en},
	number = {2},
	urldate = {2026-03-04},
	journal = {Czechoslovak Mathematical Journal},
	author = {Fiedler, Miroslav},
	year = {1973},
	pages = {298--305},
}

@article{lucas_ising_2014,
	title = {Ising formulations of many {NP} problems},
	volume = {2},
	issn = {2296-424X},
	url = {http://journal.frontiersin.org/article/10.3389/fphy.2014.00005/abstract},
	doi = {10.3389/fphy.2014.00005},
	urldate = {2026-03-04},
	journal = {Frontiers in Physics},
	author = {Lucas, Andrew},
	year = {2014},
}

@article{ponce_graph_2025,
	title = {Graph decomposition techniques for solving combinatorial optimization problems with variational quantum algorithms},
	volume = {24},
	issn = {1573-1332},
	url = {https://link.springer.com/10.1007/s11128-025-04675-z},
	doi = {10.1007/s11128-025-04675-z},
	language = {en},
	number = {2},
	urldate = {2026-03-04},
	journal = {Quantum Information Processing},
	author = {Ponce, Moises and Herrman, Rebekah and Lotshaw, Phillip C. and Powers, Sarah and Siopsis, George and Humble, Travis and Ostrowski, James},
	month = feb,
	year = {2025},
	pages = {60},
}

@article{zhou_qaoa--qaoa_2023,
	title = {{QAOA}-in-{QAOA}: {Solving} {Large}-{Scale} {MaxCut} {Problems} on {Small} {Quantum} {Machines}},
	volume = {19},
	issn = {2331-7019},
	shorttitle = {{QAOA}-in-{QAOA}},
	url = {https://link.aps.org/doi/10.1103/PhysRevApplied.19.024027},
	doi = {10.1103/PhysRevApplied.19.024027},
	language = {en},
	number = {2},
	urldate = {2026-03-04},
	journal = {Physical Review Applied},
	author = {Zhou, Zeqiao and Du, Yuxuan and Tian, Xinmei and Tao, Dacheng},
	month = feb,
	year = {2023},
	pages = {024027},
}

@article{li_large-scale_2023,
	title = {Large-{Scale} {Quantum} {Approximate} {Optimization} via {Divide}-and-{Conquer}},
	volume = {42},
	copyright = {https://ieeexplore.ieee.org/Xplorehelp/downloads/license-information/IEEE.html},
	issn = {0278-0070, 1937-4151},
	url = {https://ieeexplore.ieee.org/document/9911690/},
	doi = {10.1109/TCAD.2022.3212196},
	number = {6},
	urldate = {2026-03-04},
	journal = {IEEE Transactions on Computer-Aided Design of Integrated Circuits and Systems},
	author = {Li, Junde and Alam, Mahabubul and Ghosh, Swaroop},
	month = jun,
	year = {2023},
	pages = {1852--1860},
}

@article{esau_taiwo_comparative_2020,
	title = {{COMPARATIVE} {STUDY} {OF} {TWO} {DIVIDE} {AND} {CONQUER} {SORTING} {ALGORITHMS}: {QUICKSORT} {AND} {MERGESORT}},
	volume = {171},
	issn = {18770509},
	shorttitle = {{COMPARATIVE} {STUDY} {OF} {TWO} {DIVIDE} {AND} {CONQUER} {SORTING} {ALGORITHMS}},
	url = {https://linkinghub.elsevier.com/retrieve/pii/S1877050920312667},
	doi = {10.1016/j.procs.2020.04.274},
	language = {en},
	urldate = {2026-03-04},
	journal = {Procedia Computer Science},
	author = {Esau Taiwo, Oladipupo and Christianah, Abikoye Oluwakemi and Oluwatobi, Akande Noah and Aderonke, Kayode Anthonia and Kehinde, Adeniyi Jide},
	year = {2020},
	pages = {2532--2540},
}

@article{marchioli_quantum_2025,
	title = {Quantum {Optimization} for {Closed}-{Loop} {Scheduling} of {Earth} {Observation} {Satellite} {Formation}},
	volume = {6},
	issn = {2661-8907},
	url = {https://link.springer.com/10.1007/s42979-025-04252-2},
	doi = {10.1007/s42979-025-04252-2},
	language = {en},
	number = {6},
	urldate = {2026-03-04},
	journal = {SN Computer Science},
	author = {Marchioli, Vinicius and Boggio, Mattia and Volpe, Deborah and Massotti, Luca and Novara, Carlo},
	month = aug,
	year = {2025},
	pages = {739},
}

@misc{minato_two-step_2025,
	title = {Two-{Step} {QAOA}: {Enhancing} {Quantum} {Optimization} by {Decomposing} {K}-hot {Constraints} in {QUBO} {Formulations}},
	shorttitle = {Two-{Step} {QAOA}},
	url = {http://arxiv.org/abs/2408.05383},
	doi = {10.48550/arXiv.2408.05383},
	urldate = {2026-03-04},
	publisher = {arXiv},
	author = {Minato, Yuichiro},
	month = feb,
	year = {2025},
	note = {arXiv:2408.05383 [quant-ph]},
}

@article{giraldo-quintero_using_2022,
	title = {Using quantum computing to solve the maximal covering location problem},
	volume = {2},
	issn = {2730-6852},
	url = {https://link.springer.com/10.1007/s43762-022-00070-x},
	doi = {10.1007/s43762-022-00070-x},
	language = {en},
	number = {1},
	urldate = {2026-03-04},
	journal = {Computational Urban Science},
	author = {Giraldo-Quintero, Alejandro and Lalinde-Pulido, Juan G. and Duque, Juan C. and Sierra-Sosa, Daniel},
	month = dec,
	year = {2022},
	pages = {43},
}

@misc{farhi_quantum_2014,
	title = {A {Quantum} {Approximate} {Optimization} {Algorithm}},
	url = {http://arxiv.org/abs/1411.4028},
	doi = {10.48550/arXiv.1411.4028},
	urldate = {2026-03-04},
	publisher = {arXiv},
	author = {Farhi, Edward and Goldstone, Jeffrey and Gutmann, Sam},
	month = nov,
	year = {2014},
	note = {arXiv:1411.4028 [quant-ph]},
}

@article{hong_review_2025,
	title = {A {Review} on {Quantum} {Machine} {Learning} in {Applied} {Systems} and {Engineering}},
	volume = {13},
	copyright = {https://creativecommons.org/licenses/by/4.0/legalcode},
	issn = {2169-3536},
	url = {https://ieeexplore.ieee.org/document/11126025/},
	doi = {10.1109/ACCESS.2025.3599147},
	urldate = {2026-03-04},
	journal = {IEEE Access},
	author = {Hong, Ying-Yi and Josh Domingo Lopez, Dylan},
	year = {2025},
	pages = {144607--144631},
}

@article{ko_satellite_2026,
	title = {Satellite {Constellation} {Design} for {Minimum} {Worst}-{Case} {Revisit} {Time}},
	volume = {27},
	issn = {2093-274X, 2093-2480},
	url = {https://link.springer.com/10.1007/s42405-025-00985-9},
	doi = {10.1007/s42405-025-00985-9},
	language = {en},
	number = {1},
	urldate = {2026-03-04},
	journal = {International Journal of Aeronautical and Space Sciences},
	author = {Ko, Jaeyoul and Gwon, Beomjin and Ahn, Jaemyung},
	month = jan,
	year = {2026},
	pages = {916--929},
}

@article{imoto_optimal_2023,
	title = {Optimal constellation design based on satellite ground tracks for {Earth} observation missions},
	volume = {207},
	issn = {00945765},
	url = {https://linkinghub.elsevier.com/retrieve/pii/S0094576523001054},
	doi = {10.1016/j.actaastro.2023.02.040},
	language = {en},
	urldate = {2026-03-04},
	journal = {Acta Astronautica},
	author = {Imoto, Yuta and Satoh, Satoshi and Obata, Toshihiro and Yamada, Katsuhiko},
	month = jun,
	year = {2023},
	pages = {1--9},
}

@article{lee_regional_2023,
	title = {Regional {Constellation} {Reconfiguration} {Problem}: {Integer} {Linear} {Programming} {Formulation} and {Lagrangian} {Heuristic} {Method}},
	volume = {60},
	issn = {0022-4650, 1533-6794},
	shorttitle = {Regional {Constellation} {Reconfiguration} {Problem}},
	url = {https://arc.aiaa.org/doi/10.2514/1.A35685},
	doi = {10.2514/1.A35685},
	language = {en},
	number = {6},
	urldate = {2026-03-04},
	journal = {Journal of Spacecraft and Rockets},
	author = {Lee, Hang Woon and Ho, Koki},
	month = nov,
	year = {2023},
	pages = {1828--1845},
}

@article{correa_decomposition_2009,
	title = {A decomposition approach for the probabilistic maximal covering location-allocation problem},
	volume = {36},
	copyright = {https://www.elsevier.com/tdm/userlicense/1.0/},
	issn = {03050548},
	url = {https://linkinghub.elsevier.com/retrieve/pii/S0305054808002475},
	doi = {10.1016/j.cor.2008.11.015},
	language = {en},
	number = {10},
	urldate = {2026-03-04},
	journal = {Computers \& Operations Research},
	author = {Corrêa, Francisco De Assis and Lorena, Luiz Antonio Nogueira and Ribeiro, Glaydston Mattos},
	month = oct,
	year = {2009},
	pages = {2729--2739},
}

@article{church_maximal_1974,
	title = {{THE} {MAXIMAL} {COVERING} {LOCATION} {PROBLEM}},
	volume = {32},
	issn = {10568190},
	url = {https://linkinghub.elsevier.com/retrieve/pii/S1056819023021395},
	doi = {10.1111/j.1435-5597.1974.tb00902.x},
	language = {en},
	number = {1},
	urldate = {2026-03-04},
	journal = {Papers in Regional Science},
	author = {Church, Richard and Velle, Charles R.},
	month = jan,
	year = {1974},
	pages = {101--118},
}

\EOD

\end{document}